\documentclass [journal=ancac3,manuscript=article]{achemso}  
\usepackage[english]{babel}
\usepackage{graphicx}
\usepackage{amsmath}
\usepackage{color}
\usepackage{ulem}
\usepackage{xcolor}
\graphicspath{{Figures/}}

\let\oldAA\AA
\renewcommand{\AA}{\text{\normalfont\oldAA}}

\title{Plasmonic enhancement of aligned semiconducting graphene nanoribbons}

\author{M.~Pfeiffer}
\affiliation{Department f\"ur Chemie, Universit\"at zu K\"oln, Luxemburger Strasse 116, 50939 K\"oln, Germany}

\author{B.V.~Senkovskiy}
\affiliation{II. Physikalisches Institut, Universit\"at zu K\"oln, Z\"ulpicher Strasse 77, 50937 K\"oln, Germany}

\author{D. Haberer}
\affiliation{Department of Chemistry, University of California at Berkeley, Tan Hall 680, Berkeley, CA 94720, USA}

\author{F.R. Fischer}
\affiliation{Department of Chemistry, University of California at Berkeley, Tan Hall 680, Berkeley, CA 94720, USA}

\author{F.~Yang}
\affiliation{II. Physikalisches Institut, Universit\"at zu K\"oln, Z\"ulpicher Strasse 77, 50937 K\"oln, Germany}

\author{K.~Meerholz}
\affiliation{Department f\"ur Chemie, Universit\"at zu K\"oln, Luxemburger Strasse 116, 50939 K\"oln, Germany}

\author{Y. Ando}
\affiliation{II. Physikalisches Institut, Universit\"at zu K\"oln, Z\"ulpicher Strasse 77, 50937 K\"oln, Germany}

\author{A.~Gr\"uneis}
\affiliation{II. Physikalisches Institut, Universit\"at zu K\"oln, Z\"ulpicher Strasse 77, 50937 K\"oln, Germany}

\author{K.~Lindfors}
\affiliation{Department f\"ur Chemie, Universit\"at zu K\"oln, Luxemburger Strasse 116, 50939 K\"oln, Germany}
\email{klas.lindfors@uni-koeln.de}

\begin{document}

\begin{abstract}
We couple photoluminescent semiconducting 7-atom wide armchair edge graphene nanoribbons to plasmonic nanoantenna arrays and demonstrate an enhancement of the photoluminescence and Raman scattering intensity of the nanoribbons by more than one order of magnitude. The increase in signal allows us to study Raman spectra with high signal-to-noise ratio. Using plasmonic enhancement we are able to detect the  off-resonant Raman signals from the modified radial breathing-like mode (RBLM) due to physisorbed molecules, the 3rd order RBLM, and C-H vibrations. We find excellent agreement between data and simulations describing the spectral dependence of the enhancement and modifications of the polarization anisotropy. The strong field gradients in the optical near-field further allow us to probe the subwavelength coherence properties of the phonon modes in the nanoribbons. We theoretically model this considering a finite coherence length along the GNR direction. Our results allow estimating the coherence length in graphene nanoribbons. 
\end{abstract}

Optical nanoantennas are plasmon resonant nanoparticles that enable converting propagating light fields into localized energy and vice versa~\cite{Muehlschlegel2005,novotny:2011}. Optical antennas have found applications particularly in enhancing the light emission from subwavelength sources such as single molecules or quantum dots~\cite{Farahani:2005,Kuehn2006,Anger2006,Akselrod2014a,Kinkhabwala2009,Pfeiffer2010}, in modifying the radiation pattern~\cite{Curto:2010,Dregely:2014,Jun2011}, and in enhancing spectroscopic signals~\cite{Kuehn2006,Anger2006,Hartschuh2005,Wokaun1983,Stockle2000}. For low quantum yield emitters optical antennas can be used to significantly increase the quantum yield by enhancing the radiative decay rate~\cite{Hartschuh2005,Wokaun1983}. Extreme field gradients near the nanoscale plasmonic structures even allow modifying the selection rules for optical transitions~\cite{Takase2013,Jain2012,Iida2009}. Plasmonic nanoantennas thus open up new domains in spectroscopy for weakly luminescent materials.

Graphene nanoribbons (GNRs) have attracted a lot of interest in recent years due to their remarkable electronic and optical properties~\cite{Nakada1996,Ruffieux_NatCom14,Chernov2013,Lim2015,louie07-excitonic,prezzi08-manybody,Kim-NatNan13,Stampfer2009,Li2008,Son2006,Han2007,Senkovskiy17}. Various species of GNRs can selectively be grown with atomic precision and with a high degree of alignment using bottom-up fabrication~\cite{Chen2013,muellen10-ribbon,Ruffieux_ZigZag,ADMA201505738}. Even atomically precise doping and heterostructures can be realized in this way~\cite{Chen15-heterojunction,B-GNRs,Fischer_JACS_15,Carbonell-Sanroma2017,Nguyen2016}. The properties of GNRs are strongly influenced by ribbon width and geometry of their edges~\cite{Son2006,Wang2016}. Here we focus on seven atomwide armchair GNRs (7-AGNRs), which are semiconducting. Recently, photoluminescence~\cite{Chernov2013,Senkovskiy17} and electroluminescence~\cite{Chong2017} from armchair edge nanoribbons (AGNRs) was observed. Armchair edge graphene nanoribbons exhibit a broad fluorescence spectrum centered in the red-spectral domain, which for pristine nanoribbons is weak in intensity~\cite{Senkovskiy17,Zhao2017}. The low quantum yield is attributed to dark excitons degenerate with the optically active state~\cite{Yang2007,Prezzi2007}. It was recently shown that the degeneracy can be removed by introducing defects into the ribbon~\cite{Senkovskiy17,Zhao2017_SciDir}. To avoid deterioration of the electronic properties of GNRs due to a high density of defects, we apply plasmonic nanoantennas to further increase the quantum efficiency of GNRs.  By placing the nanoribbons in the near-field of arrays of plasmon resonant gold nanoparticles we increase both the excitation and radiative decay rates. This results in more than an order of magnitude increase in the Raman scattering and photoluminescence rate. The strong field gradients of the optical near field additionally result in modified Raman scattering and allow us to probe the phonon correlation length.

\section{Results and discussion}
\begin{figure} [t!]
	\centering
		\includegraphics{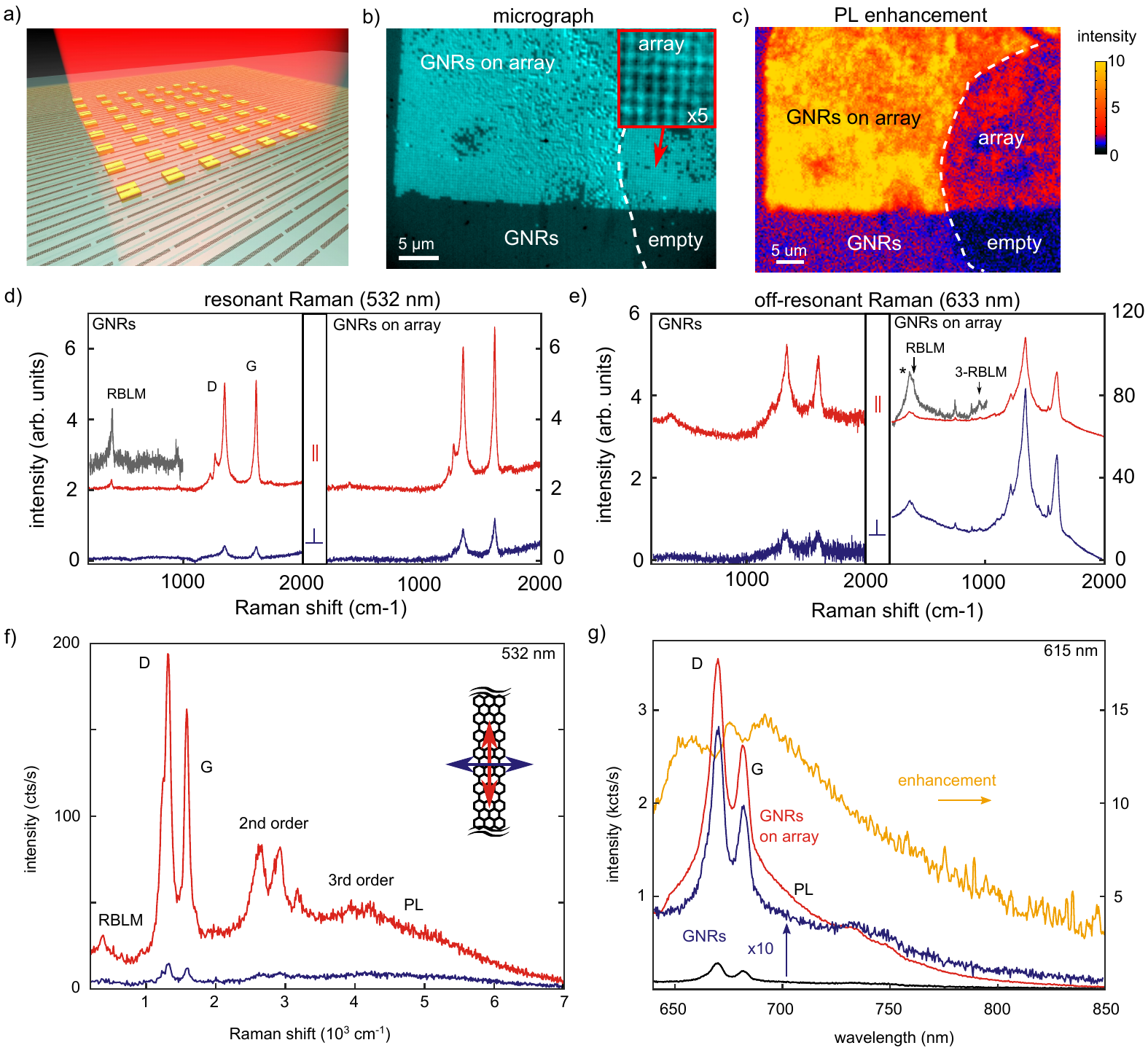}
	\caption{a) Sketch of GNRs on a plasmonic nanoantenna array. b) The region of the plasmonic array covered with nanoribbons is clearly visible in a reflection micrograph under white light illumination. c) Photoluminescence micrograph shows that photoluminescence (PL) and Raman scattering (D- and G- like modes) from the same sample region as in (b) are increased by an order of magnitude. d) Raman spectra for aligned GNRs for 532~nm and e) 633~nm excitation wavelength with incident polarization along (red) and perpendicular (blue) to the ribbons. Spectra for GNRs and GNRs on array are shown in the left and right half, respectively. From the magnified acoustic region (gray), we observe a second peak (*) for the RBLM in the off-resonant Raman spectrum. Spectra for polarization along ribbon for 532~nm are offseted by 2, spectra for 633~nm by 3 and 60, for GNRs and GNRs on array, respectively. f) Polarization dependence of the photoluminescence of GNRs (excitation wavelength 532~nm). g) Spectra for GNRs (black, 10 times magnification blue) and GNRs on array (red). The spectral enhancement (yellow) is the ratio of the spectrum on the array and besides it.}\label{fig:fig1}
\end{figure}

We synthesized densely lying aligned 7-AGNRs on a Au(788) surface by on-surface assembly from molecular precursors. The obtained single layer of GNRs is transferred by so-called "bubbling" technique onto arrays of plasmonic antennas on a glass substrate with preserving the ribbon alignment~\cite{Senkovskiy17,Gao2012,Wang2011}.
The ribbons alignment direction is along one of the arrays' main axes as schematically illustrated in Fig.~\ref{fig:fig1}a. The plasmonic structures are square arrays of square and cylindrical disk gold nanoantennas. We vary the period of the arrays to tune the plasmon resonance of the lattice. Figure~\ref{fig:fig1}b shows an optical micrograph of a section of an array with transferred ribbons. The array shown here consists of square-shaped nanoantennas with a nominal side length of 100~nm and a period of 350~nm. The spectral position of the surface plasmon resonance of the individual array element is approximately 680~nm based on simulations (see Fig.~\ref{fig:figS1}a). The edge of the nanoribbon coverage is indicated by the dashed white line. The individual array elements can be resolved in the magnified image shown in the inset of Fig.~\ref{fig:fig1}b.

We first consider arrays with periods smaller than the emission wavelength in the substrate (and thus also superstrate). In this case there is no coupling of the exciton in the nanoribbon to long-range grating plasmons. Figure~\ref{fig:fig1}c shows a micrograph of the spectrally integrated emission or scattering consisting of spectrally broad photoluminescence and Raman scattering for the sample region shown in Fig.~\ref{fig:fig1}b. Here the excitation wavelength is 595~nm and light with wavelengths longer than 645~nm are collected. We identify four regions with different brightness levels. These regions are in ascending order of brightness the clean substrate (empty), regions beside the plasmonic nanostructures covered by a homogeneous layer of nanoribbons (GNRs), regions where the array is not covered by GNRs (array), and the array covered with a homogeneous layer of GNRs (GNRs on array). From the comparison of regions (GNRs) and (GNRs on array) we can deduce a (spatially averaged) brightness increase of about one order of magnitude due to coupling to the array. The collected light is characterized by measuring spectra in these four regions. The presence of characteristic Raman features, which originate from nanoribbons, in regions (GNRs) and (GNRs on array), and their absence in regions (empty) and (array) confirms the attribution of the sample regions. Figure~\ref{fig:fig1}d shows the spectrum for GNRs on and next to the array for 532~nm excitation wavelength. The spectra from the uncovered regions on and besides the array were used for background subtraction.\\

\textbf{Raman scattering}\\

\begin{figure}[t!]
\centering
\includegraphics{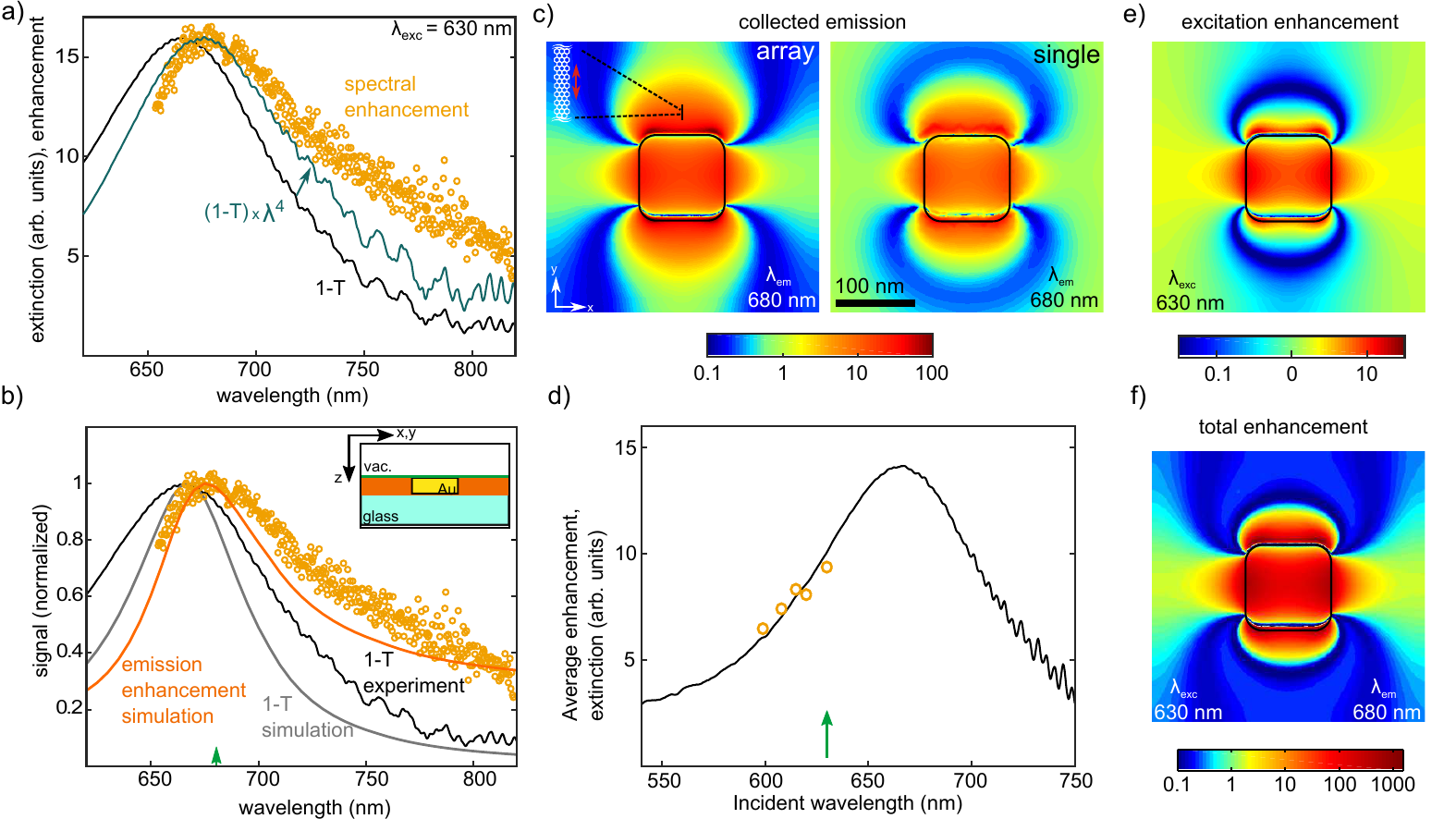}
\caption{a) The comparison of spectral enhancement (orange) and the extinction spectrum (black) of the GNR-covered array shows a spectral shift, which is explained by the deviation between near-field enhancement and far-field spectrum. The spectral shape of the near-field enhancement can be reproduced around the peak from the extinction spectrum by multiplication with the factor $\lambda^{4}$ (dark green). b) Finite element simulations are in good agreement with experiment for enhancement and extinction spectra (red and gray, respectively). The emission enhancement is obtained by spatial averaging in a layer above the substrate (orange domain in the inset). c) The enhancement of collected emission from emitters in a plane above the array and above an isolated nanoantenna at the resonance wavelength [marked in panel (b) by a green arrow]. The regions of suppression are due to destructive interference between direct and array coupled emission pathways. d) The spectrally averaged enhancement as a function of the excitation wavelength (orange circles) follows the extinction spectrum (green solid line). e) Excitation enhancement in the unit cell of the array for the illumination wavelength 630~nm [marked with an arrow in panel (d)]. f) The total enhancement at the peak of the enhancement for illumination (630~nm) and emission (680~nm). In the hot spots enhancement factors of up to 2000 are obtained.}\label{fig:fig2}
\end{figure}

Figure~\ref{fig:fig1}d and e display the Raman spectra for 532~nm (Fig.~\ref{fig:fig1}d) and 633~nm (Fig.~\ref{fig:fig1}e) excitation wavelength for polarization along and perpendicular to the alignment direction of the GNRs. The measurements were performed on the same spots on the sample. We observe that the enhancement for GNRs on array for 532~nm is negligible, while for 633~nm we obtain an enhancement of the signal by a factor of approximately 20 for light polarized along the ribbon. We remark that the enhancement factor represents a spatial average over the unit cell of the array. Within the unit cell we expect there to exist hot spots around the antenna element where the signal is enhanced much more than on average. The observed wavelength dependence is consistent with the extinction spectrum of the plasmonic array (see Fig.~\ref{fig:fig2}a). The 532~nm is far away from the plasmon resonance and no significant enhancement is expected.

Surprisingly, for 633~nm wavelength incident light we observe that the Raman signal on the plasmonic array is stronger for incident light polarized perpendicular to the ribbon. In the absence of the plasmonic nanostructure the scattering for polarization along the ribbon is significantly stronger than for the orthogonal polarization as evidenced by the Raman spectra collected besides the array and reported earlier~\cite{Senkovskiy17}. The enhancement is in this case approximately a factor of 90. Based on additional polarization-resolved measurements (see Fig.~\ref{fig:figS3} in Supporting Information), where the polarization state of the emission is analyzed, we conclude, that the observed polarization dependence of the enhancement is predominantly explained by a change of the polarization state of the incident light in the vicinity of the plasmonic nanoantennas. The scattered light is in this case still strongly polarized along the ribbons. Using finite element simulations (see below for details about simulations) we find, that for perpendicular polarization (electric field along $x$-axis, see Fig.~\ref{fig:fig2}c), there are hotspots in the near field with a noticeable enhancement of the local field along the ribbons. These hotspots of the incident field overlap with the regions of enhanced scattering of the $y$-oriented dipole at the Raman scattering wavelength. Due to this we observe the change of the polarization anisotropy (see Supporting Information for further details).
Considering the Raman  modes for GNRs on the array (Fig.~\ref{fig:fig1}d and e), we observe an enhancement of the radial breathing-like mode (RBLM), which, due to its resonant nature, is not typically well visible with high signal-to-noise ratio for an illumination wavelength of 633~nm. Instead we observe a broadened peak (in comparison to the spectrum obtained with 532~nm excitation), which can be resolved as two peaks in the plasmonically enhanced spectrum. Fitting the spectrum with two Lorentzian functions, assuming the RBLM of pristine GNRs at 396.0~cm$^{-1}$, we obtain a sideband at 363.9~cm$^{-1}$ (see Supporting Information, Fig.~\ref{fig:figS4}a). The shift to smaller wavenumbers indicates a broader effective width of the GNRs. As the polarization anisotropy of the sideband is not modfied we attribute it to the attachment of atoms or small molecules, e.g. oxidization at the GNRs edges~\cite{Verzhbitskiy16,Narita2014}.
Besides the two narrow bands at 750~cm$^{-1}$ and 889~cm$^{-1}$, which we identify as C-H modes~\cite{Gillen2009}, we observe a broader peak at 953~cm$^{-1}$ (see Fig.~\ref{fig:figS4}c of the Supporting Information). This mode is attributed to the next odd higher-order transverse acoustic 3-RBLM mode as theoretically predicted\cite{Gillen2010,Ma2017}. 

\textbf{Photoluminescence}\\
We now move to enhancement of photoluminescence (PL). In Fig.~\ref{fig:fig1}f is shown the emission spectrum on a broader spectral range than in the Raman spectroscopy experiments. We observe a broad emission peak at 700~nm, which is the photoluminescence of GNRs\cite{Senkovskiy17}. On top of the broad luminescence feature we identify the second and third order of the D-like and G-like peaks and their combinations. From polarization resolved measurements on reference areas without plasmonic structures (see Fig.~\ref{fig:fig1}f) we find for the PL a similar polarization anisotropy as for the Raman peaks. In Fig.~\ref{fig:fig1}g we show the emission spectra for GNRs and GNRs on the array (black and red, respectively) for an excitation wavelength of 615~nm. Here we identify the PL as a broad background with the D- and G-like Raman bands at $\approx$~670~nm and $\approx$~680~nm, respectively. The enhancement due to the plasmonic array is obtained as the ratio of the spectra on and away from the array (yellow). A maximum enhancement factor of more than 14 occurs at a wavelength of around 670~nm. Here we note, that we observe dips in the enhancement at the spectral positions of the Raman bands. This is due to the different coherence properties of photoluminescence and Raman scattering \cite{Cancado2014,Beams2014} and is discussed in more detail below.

We now want to relate the spectral enhancement (Fig.~\ref{fig:fig1}g) to the optical properties of the antenna array. Due to the large size of the single array elements, scattering dominates in comparison to absorption (see Fig.~\ref{fig:figS1}). The extinction [defined here as 1-transmission (1-T)] spectrum hence reflects the scattering properties. Figure~\ref{fig:fig2}a shows the extinction spectrum for the antenna array from Fig.~\ref{fig:fig1}b (period 350~nm) together with the spectral enhancement for an illumination wavelength of 630~nm. The total enhancement has a maximum value of 16 at a wavelength of 680~nm. This is approximately 13~nm red-shifted from the spectral position of the peak of the extinction spectrum. This shift can be explained as the spectral deviation of the maximum enhancement in the near-field from the plasmon resonance observed in far-field scattering~\cite{Messinger1981,Alonso-Gonzalez2013,Zuloaga2011}. A simple electric point-dipole model can account for this shift when we note the difference in the wavelength dependence of the intensity of the near-field $I_\mathrm{NF}(\lambda)$ and the far-field $I_\mathrm{FF}(\lambda)$~\cite{Hecht2006} of a dipole. We obtain
\begin{equation}
I_\mathrm{NF}(\lambda)/I_\mathrm{FF}(\lambda) \propto \lambda^4,
\end{equation}
where $\lambda$ is the wavelength of light. By taking into account also the excitation process (see Supporting Information) we obtain the near-field spectrum shown in Fig.~\ref{fig:fig2}a (dark green). It matches well the experimentally obtained spectral enhancement around the peak (approximately 650 to 700~nm). At longer wavelengths we experimentally observe a larger enhancement than predicted by the single dipole approximation. As the grating period becomes smaller compared to the wavelength we have to include interactions between neighbouring array elements via intermediate- and far-field components. This results in a deviation from the simple model introduced above.

For a quantitative comparison between experiment and model we use finite element simulations to calculate the increase in the collected emission from the GNRs. Here we apply the Lorentz reciprocity theorem to deduce the modification in far-field fluorescence due to a local source near the plasmonic structure~\cite{Janssen2010,Zhang2015,Chen:15,Pfeiffer2014,Koenderink17}.
The nanoribbons are modeled as electric point dipoles oriented along the $y$-axis (see Fig.~\ref{fig:fig2}c) and we consider here emission only along the surface normal direction. Due to the absence of long-range grating plasmons the angular distribution of emission is not strongly modified due to the array. The spatially averaged emission enhancement in the region of the GNRs (orange domain in the inset of Fig.~\ref{fig:fig2}b) is shown in Fig.~\ref{fig:fig2}b for the emission wavelength at the plasmon resonance. In our simulations we assume that the lateral size of the nanoantennas is increased by 10~$\%$ from the nominal dimensions and the corners are rounded with a radius of 20~nm. This is in agreement with our experience from comparing sizes and shapes of nominal and fabricated structures using electron beam lithography. See supporting information for full details about the simulations. Calculated and measured extinction spectra are in good agreement (see Fig.~\ref{fig:fig2}b) justifying the increased size of the plasmonic nanoantennas. The scattering resonance of the isolated antenna element obtained from separate calculations is at 691~nm (see supporting information). The simulated spectrum reproduces the elevated enhancement values at longer wavelengths compared to the simple dipole model. From the simulation we extract the spatially averaged increase of the emission due to the array. At the peak of the spectral enhancement we obtain an emission enhancement factor of 5.5.

We now turn to the spatial distribution of the emission enhancement around the plasmonic nanoantennas. Figure~\ref{fig:fig2}c shows the simulated enhancement in a plane 2~nm above the array (green layer in inset of Fig.~\ref{fig:fig2}b) for the plasmon resonance wavelength (680~nm). Additionally, for comparison the same quantity is displayed for a single isolated antenna. We observe that on resonance the enhancement reaches values of approximately 100 but that there are also regions of suppressed emission. Most surprisingly, the emitters on top of the plasmonic nanoantennas are strongly enhanced. For an emitter oriented along the metal surface the induced image dipole $\mathbf{p}_\mathrm{ind}$ is out-of-phase with the emitting dipole $\mathbf{p}$ and one expects suppressed emission due to destructive interference~\cite{Hecht2006}. Here the increased radiation is a result of the collected emission coming dominantly via the antenna. This can be qualitatively described by treating the antenna as a polarizable oblate ellipsoid. The ratio of the total dipole moment with and without antenna can be expressed as
\begin{equation}\label{eq:2}
\frac{\left| \mathbf{p} + \mathbf{p}_\mathrm{ind} \right|^2}{\left| \mathbf{p} \right|^2}=
\left| 1 - \frac{1}{4 \pi \epsilon_0}{\alpha}_\mathrm{yy}\frac{\exp(ikR)}{R^3} \right|^2,
\end{equation}
where $\alpha_{yy}$ is the relevant component of the polarizability tensor of the nanoantenna, $R$ is the distance between the antenna and emitter dipoles, and $k=2\pi/\lambda$. Due to the geometry here only the transverse component of the emitter's near-field is relevant. The induced and primary dipoles are indeed out of phase (negative sign in Eq.~\ref{eq:2}). However, the resonantly induced dipole in the antenna significantly exceeds the primary dipole, resulting in enhancement. We remark here that for a spherical antenna or a continuous metal film the enhancement turns to a suppression of emission as intuitively expected.

The wavelength dependence of the excitation process is shown in Fig.~\ref{fig:fig2}d. Here the spectrally averaged enhancement (see Fig.~\ref{fig:fig2}a for spectral enhancement) for the array considered in Fig.~\ref{fig:fig2}a--c is shown as a function of the excitation wavelength. We point out that the averaged enhancement is approximately 50~$\%$ of the maximum enhancement, which occurs at a wavelength of about 680~nm.
The enhancement as a function of excitation wavelength closely follows the far-field resonance (extinction spectra). To disentangle the enhancement of the excitation and emission processes we simulate the intensity distribution in the unit cell of the array for one of the excitation wavelengths (630~nm) used in the experiments. As seen in Fig.~\ref{fig:fig2}e, the largest enhancement region is localized on top of the antenna. The total enhancement pattern that is probed in the experiment is given by the product of excitation and emission enhancement. This is shown in Fig.~\ref{fig:fig2}f for the excitation wavelength wavelength (630~nm) marked in Fig.~\ref{fig:fig2}d and for emission at 680~nm wavelength (green arrow in Fig.~\ref{fig:fig2}b), i.e. at the maximum of the spectral enhancement. The overlap of the regions of enhancement in emission and excitation occur only in regions above the antenna or very close to it. Double-resonant antennas could thus be very useful to match the regions of enhancement~\cite{Thyagarajan2012}. From the simulated field distributions of Fig.~\ref{fig:fig2}f a spatially averaged enhancement factor of 21.0 and a maximum of about 2000 close to the antenna is obtained. For comparison, the maximum average enhancement for Raman scattering is about 20 for this wavelength (see Fig.\ref{fig:fig1}e). The average enhancement is thus in good agreement with the experimental data.

\begin{figure}[!t]
\centering
\includegraphics{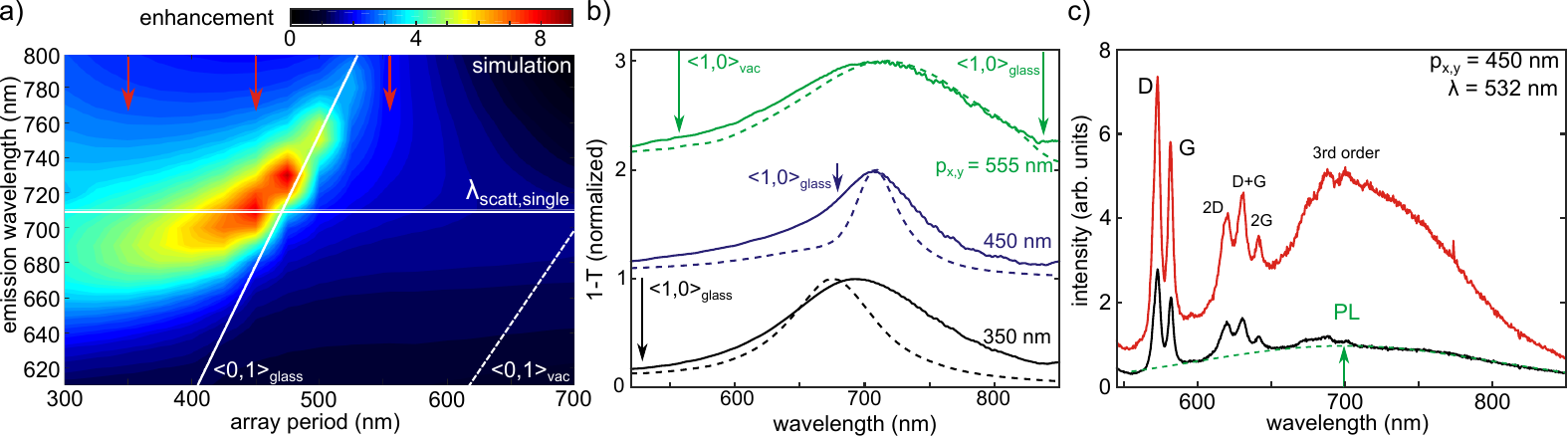}
\caption{a) The emission enhancement as a function of array period and wavelength displays a maximum close to the point where the grating order intersects the plasmon resonance wavelength. b) Extinction spectra for arrays with periods of 350~nm, 450~nm, and 555~nm, obtained experimentally (solid lines) and from simulations (dashed lines). The arrows indicate grating orders. The spectra have been shifted vertically for clarity. c) Photoluminescence spectra for GNRs on the array (red) and from a reference region (black) for illumination with 532~nm light and polarization along the ribbon orientation. The green dotted line shows the fitted contribution from photoluminescence. 
}\label{fig:fig3}
\end{figure}

After studying subwavelength period arrays we now move to exploring the influence of the grating period on the enhancement of the light emission from nanoribbons. We keep the dimensions and shape of the array elements constant. Here we use arrays of round gold disks with a radius of 70~nm and a height of 30~nm. From simulations we obtain the surface plasmon resonance of the isolated nanoantennas at 707~nm (see Fig.~\ref{fig:figS1}b). The critical period $\Lambda_C$ corresponding to the first grating order into the substrate half space when illuminated with light of the wavelength $\lambda_0$ at normal incidence can be calculated as~\cite{Felidj2005}
\begin{equation}
\Lambda_C = \frac{\lambda_{0}}{n_\mathrm{sub}},
\end{equation}
where $n_\mathrm{sub}$ is the refractive index of the substrate. For a resonance wavelength of 707~nm and a refractive index of 1.51 of the glass substrate the critical period is 468~nm.

We start by investigating the enhancement in the unit cell for arrays with different periods using simulations. We vary the array period and calculate the average intensity enhancement in a 32~nm thick domain above the substrate. The procedure is the same as in the case of subwavelength period arrays considered in Fig.~\ref{fig:fig2}b. Figure~\ref{fig:fig3}a shows the calculated enhancement as a function of wavelength and array period. The array modes into the glass substrate and air superstrate are indicated as solid and dashed white lines, respectively. The calculated plasmon resonance of the single antenna, which is close to the emission wavelength of the GNRs, is indicated with a horizontal white line. The calculated enhancement reaches a maximum of about 8 on the left side of the crossing between array and plasmonic mode. This occurs for a period of 450~nm at a wavelength of about 710~nm. According to these calculations we fabricated three samples with the optimized period of 450~nm and for comparison with periods of 350~nm and 555~nm. The red arrows in Fig.~\ref{fig:fig3}a indicate these periods.

The extinction spectra for the three arrays are shown in Fig.~\ref{fig:fig3}b. The solid lines indicate the measured extinction spectra while the dashed curves are extinction spectra obtained from finite element simulations. For the investigated grating periods the array resonances shifts from shorter wavelengths to close to the plasmon resonance for the structure with a period of 450~nm, and then to longer wavelengths for a period of 555~nm. The grating orders into superstrate and substrate are indicated by vertical arrows. When the period is increased from below the critical period to a value close to the critical period, we observe a red-shift of the spectral position of the resonance. The resonance for the array with the 350~nm period is more narrow than the single particle resonance (see Fig.~\ref{fig:figS1}) due to the reduced radiation damping when no grating orders are present. The width becomes even more narrow and the spectral shape becomes asymmetric when the Rayleigh anomaly approaches the plasmon resonance from the short wavelength side~\cite{Christ2005}. The broadest resonance is observed, when the plasmon resonance is at a shorter wavelength than the first diffraction order into the substrate half-space. This is due to the mutual coupling of the elements and a simultaneous shortening of the plasmon lifetime~\cite{Lamprecht2000,Auguie2010}. The absence of clear grating anomalies in the transmission spectra is explained by the refractive index mismatch between substrate and superstrata~\cite{Felidj2005,Hicks2005}.

In the photoluminescence experiments we focus on the array with a period of 450~nm where we expect the maximum enhancement. Figure~\ref{fig:fig3}c shows the measured photoluminescence spectra from GNRs on and besides the array. To be able to study modifications of the emission all measurements were carried out with an excitation wavelength of 532~nm. For this wavelength gold shows no plasmonic response. This is confirmed by simulations, which show negligible spatially averaged excitation enhancement for this wavelength. A Gaussian fit to the PL component of the spectrum from nanoribbons besides the array is shown with green dashed line. At the enhancement maximum of 710~nm our measurements show a maximum enhancement factor of approximately 5. This is in good agreement with the enhancement factor of 6.3 obtained with simulations. We remark that for this excitation wavelength the plasmonic nanostructure actually results in a reduced excitation efficiency.

\textbf{Difference in enhancement of Raman scattering and photoluminescence}

We now return to the different enhancement obtained for photoluminescence and Raman scattering. The illumination and the detection wavelengths are the same for both processes. As photoluminescence at room temperature for the nanoribbons is a purely incoherent process, the total emission is equal to the incoherent sum of the radiation from the individual emitters. The total (measured) enhancement is the product of the enhancements for excitation and emission, as discussed above. For plasmonic near-field enhancement of Raman scattering the process is partially coherent. This was recently predicted and experimentally confirmed for tip-enhanced Raman scattering of the D, G, and 2D Raman modes of graphene~\cite{Cancado2014,Beams2014}. In these works a phonon correlation length of about 30~nm explained the experimental results. The partially coherent nature of the scattering process leads to a difference of the enhancement of the Raman scattering from photoluminescence due to interference. From the experimental data presented in Fig.~\ref{fig:fig2} we can extract the contributions of photoluminescence and Raman scattering by fitting these with Gaussian functions. From this we obtain enhancement factors of 11.8 and 13.8
for Raman scattering and photoluminescence, respectively. As the ribbon length is approximately 30~nm the reduction of the Raman enhancement could be due to coherence effects on this scale. For a propagating electromagnetic wave this would not be observable. However, for evanescent fields around optical antennas the phase changes on a much shorter length scale than the wavelength (see Fig.~\ref{fig:fig2}c). This enables destructive interference for scattering from points with separations shorter than the coherence length that could result in a decrease in the enhancement for Raman scattering.

We apply the theory of Ref.~\cite{Cancado2014} to the geometry of Fig.~\ref{fig:fig2}c to study the influence of coherence on the observed enhancement. Briefly, using full field simulations at the incident and scattered wavelength $\lambda_{i}$ and $\lambda_{s}$, respectively, we calculate the average total enhancement $S(\lambda_{s})$ including a finite coherence length $L_c$ in the direction of the ribbons ($y$-direction). This is performed by introducing the one-dimensional correlation function
\begin{equation}
f_{c}(y_1,y_2)=\frac{\exp\left(- \frac{\left| y_1 - y_2\right|}{L_{c}^2}\right)}{\sqrt{\pi} L_{c}}
\end{equation}
between two points $(x_1,y_1,z_1)$ and $(x_2,y_2,z_2)$. We obtain the average enhancement signal $S(\lambda_{s})$ for $y$-polarized light following the model in Ref.~\cite{Cancado2014} as
\begin{multline}\label{eq:coherence}
S(\lambda_{s})=A\int_{A_\mathrm{unit}}\int_{A_\mathrm{unit}}dA_1dA_2
f_{c}(y_1,y_2){G}^*(\mathbf{r}_0,x_1,y_1;\lambda_{s})_{yy}{G}_{yy}(\mathbf{r}_0,x_2,y_2;\lambda_{s})\\
\times {\alpha}_{yy}{\alpha}_{yy}{E}^*_{y}(x_1,y_1;\lambda_{i}){E}_{y}(x_2,y_2;\lambda_{i}),
\end{multline}
where $A$ is a proportionality constant and $G_{yy}(\mathbf{r}_0,x_i,y_i;\lambda_{s})$ with $i=1,2$ is the $yy$-component of the dyadic Green's function evaluated at the field point $\mathbf{r}_0$, which here corresponds to the detector, and for the source point $(x_i,y_i)$. In Eq.~\ref{eq:coherence} ${E}_{y}(x_i,y_i;\lambda_{i})$ is the electric field at the position $(x_i,y_i)$ and ${\alpha}_{yy}$ is the polarizability of the ribbon. Using the reciprocity theorem, we can replace the Green's function by the numerically calculated value for the electric field when a $y$-polarized plane wave is incident on the structure. 
Comparing the reduction of 14~\% found in the experiment with the calculated reduction for the detected Raman scattering intensity from the fully incoherent situation as a function of the coherence length, we deduce a coherence length along the GNRs of 15~nm. We thus conclude that the Raman coherence length seems to mainly be limited by the length of the ribbons and not by defects in the material. 

\subsection{Conclusions}

We have demonstrated more than an order of magnitude enhancement of the Raman scattering and photoluminescence from graphene nanoribbons. The increased signal allows us to perform Raman spectroscopy even for non-resonant excitation. We observe distinct differences in the Raman spectrum obtained with resonant Raman excitation and plasmonically enhanced non-resonant illumination. For plasmonically enhanced off-resonant Raman the RBLM consists of two spectral lines. The plasmonic enhancement for Raman scattering is smaller than for photoluminescence at the same wavelength. We analyze this in terms of the coherence properties of the two processes. Our experimental data indicates that the phonon coherence length of the graphene nanoribbons is limited by the length of the ribbon and not by defects.

In order to optimize the plasmonic enhancement we analyze the spectral properties of the enhancement and the field localization in the unit cell with the help of full-wave simulations. We observe that even in the case of a delocalized array resonance the enhancement is localized to a very small fraction of the unit cell. Interestingly, we note that for the disk-shaped antennas used in this work the emission is enhanced on top of the antennas, which is not usually the case for emitters with their transition dipole moment along the metal surface. This illustrates the importance of the proper design of the antenna to enhance the light-matter interaction of atomically thin materials.

Our work shows how to enhance light-matter interaction of graphene nanoribbons on a large area using plasmonic gratings. Plasmonic enhancement allows Raman spectroscopy even using non-resonant excitation and creating field gradients on the nanoscale. This enables selective excitation of modes with different symmetry and exploring their wavelength dependence. Finally, the performance of devices such as plasmonically enhanced nanoscale photovoltaic or light emitting devices can
be boosted using plasmonic arrays.

\section{Methods}

\subsection{Array fabrication}

The plasmonic structures were designed using full-field simulations using a finite element solver (Comsol Multiphysics) and fabricated using electron beam lithography on glass substrates. The patterns of plasmonic nanoantenna arrays were exposed on a 200~nm thick double layer poly-methyl methacrylate (PMMA) resist on a glass substrate. The bottom layer (molecular weight 250000) is more sensitive than the top layer (molecular weight 950000), resulting in an undercut that facilitates lift-off processing. A 30 nm thick layer of conductive polymer (eSpacer 300Z) was spin coated on top of the resist to prevent charging of the sample. The sample was developed after exposure and 3~nm of chromium and 30~nm of gold were deposited by thermal evaporation from resistively heated sources. After a lift-off process the metal remains on the substrate only in the exposed regions.

\subsection{Growth and transfer of GNRs}

Seven atom wide armchair-edge graphene nanoribbons were grown on a single crystalline Au~(788) surface. The ribbons are aligned along the crystal edges of the gold crystal on a macroscopic scale. We transferred the ribbons from the gold crystal to the plasmonic structures using bubbling transfer as described earlier~\cite{Senkovskiy17,Gao2012,Wang2011}. The nanoribbons were placed on the plasmonic arrays with their orientation aligned along one of the arrays' main axis.

\subsection{Optical characterization}

All optical measurements were carried out in a sample-scanning micro-photoluminescence setup. The excitation was performed with a super-continuum source (NKT, SuperK EXTREME) in combination with a variable optical bandpass filter (NKT SuperK VARIA). To be able to perform tunable Raman spectroscopy, the spectral bandwidth of the incident light was further narrowed down to below 1~nm using a home-built optical filter based on a diffraction grating and slit assembly. The sample illumination was performed through a microscope objective with a numerical aperture of 0.8 (MPlanFl, Olympus). Emitted light was collected with the same objective and separated from the incident light using a beam splitter and long-pass filters. The incident laser light was weakly focused in the back focal plane of the microscope objective. With this the illumination spot size was increased to approximately 6~$\mu$m. This narrows down the angular spectrum of the illumination and improves the comparability to simulation, where we consider excitation by a plane wave (at normal incidence). All spectrally integrated signals were recorded with a single photon counting module. Spectra were acquired using a grating spectrometer equipped with a deep-cooled charge-coupled device (CCD) camera.

\section{Acknowledgments}

This work has been supported by the University of Cologne through the Institutional Strategy of the University of Cologne within the German Excellence Initiative. A.G. and B.S. acknowledge the ERC grant No. 648589 "SUPER-2D".

\bibliography{optical_save_171002}

\providecommand{\latin}[1]{#1}
\makeatletter
\providecommand{\doi}
  {\begingroup\let\do\@makeother\dospecials
  \catcode`\{=1 \catcode`\}=2\doi@aux}
\providecommand{\doi@aux}[1]{\endgroup\texttt{#1}}
\makeatother
\providecommand*\mcitethebibliography{\thebibliography}
\csname @ifundefined\endcsname{endmcitethebibliography}
  {\let\endmcitethebibliography\endthebibliography}{}
\begin{mcitethebibliography}{69}
\providecommand*\natexlab[1]{#1}
\providecommand*\mciteSetBstSublistMode[1]{}
\providecommand*\mciteSetBstMaxWidthForm[2]{}
\providecommand*\mciteBstWouldAddEndPuncttrue
  {\def\EndOfBibitem{\unskip.}}
\providecommand*\mciteBstWouldAddEndPunctfalse
  {\let\EndOfBibitem\relax}
\providecommand*\mciteSetBstMidEndSepPunct[3]{}
\providecommand*\mciteSetBstSublistLabelBeginEnd[3]{}
\providecommand*\EndOfBibitem{}
\mciteSetBstSublistMode{f}
\mciteSetBstMaxWidthForm{subitem}{(\alph{mcitesubitemcount})}
\mciteSetBstSublistLabelBeginEnd
  {\mcitemaxwidthsubitemform\space}
  {\relax}
  {\relax}

\bibitem[M{\"u}hlschlegel \latin{et~al.}(2005)M{\"u}hlschlegel, Eisler, Martin,
  Hecht, and Pohl]{Muehlschlegel2005}
M{\"u}hlschlegel,~P.; Eisler,~H.-J.; Martin,~O. J.~F.; Hecht,~B.; Pohl,~D.~W.
  Resonant Optical Antennas. \emph{Science} \textbf{2005}, \emph{308},
  1607--1609\relax
\mciteBstWouldAddEndPuncttrue
\mciteSetBstMidEndSepPunct{\mcitedefaultmidpunct}
{\mcitedefaultendpunct}{\mcitedefaultseppunct}\relax
\EndOfBibitem
\bibitem[Novotny and van Hulst(2011)Novotny, and van Hulst]{novotny:2011}
Novotny,~L.; van Hulst,~N. {Antennas for light}. \emph{Nature Photonics}
  \textbf{2011}, \emph{5}, 83--90\relax
\mciteBstWouldAddEndPuncttrue
\mciteSetBstMidEndSepPunct{\mcitedefaultmidpunct}
{\mcitedefaultendpunct}{\mcitedefaultseppunct}\relax
\EndOfBibitem
\bibitem[Farahani \latin{et~al.}(2005)Farahani, Pohl, Eisler, and
  Hecht]{Farahani:2005}
Farahani,~J.~N.; Pohl,~D.~W.; Eisler,~H.-J.; Hecht,~B. Single Quantum Dot
  Coupled to a Scanning Optical Antenna: A Tunable Superemitter. \emph{Phys.
  Rev. Lett.} \textbf{2005}, \emph{95}, 017402\relax
\mciteBstWouldAddEndPuncttrue
\mciteSetBstMidEndSepPunct{\mcitedefaultmidpunct}
{\mcitedefaultendpunct}{\mcitedefaultseppunct}\relax
\EndOfBibitem
\bibitem[K\"uhn \latin{et~al.}(2006)K\"uhn, Hakanson, Rogobete, and
  Sandoghar]{Kuehn2006}
K\"uhn,~S.; Hakanson,~U.; Rogobete,~L.; Sandoghar,~V. Enhancement of
  single-molecule fluorescence using a gold nanoparticle as an optical
  nanoantenna. \emph{Phys. Rev. Lett.} \textbf{2006}, \emph{97}, 017402\relax
\mciteBstWouldAddEndPuncttrue
\mciteSetBstMidEndSepPunct{\mcitedefaultmidpunct}
{\mcitedefaultendpunct}{\mcitedefaultseppunct}\relax
\EndOfBibitem
\bibitem[Anger \latin{et~al.}(2006)Anger, Bharadwaj, and Novotny]{Anger2006}
Anger,~P.; Bharadwaj,~P.; Novotny,~L. Enhancement and quenching of
  single-molecule fluorescence. \emph{Phys. Rev. Lett.} \textbf{2006},
  \emph{96}, 11302\relax
\mciteBstWouldAddEndPuncttrue
\mciteSetBstMidEndSepPunct{\mcitedefaultmidpunct}
{\mcitedefaultendpunct}{\mcitedefaultseppunct}\relax
\EndOfBibitem
\bibitem[Akselrod \latin{et~al.}(2014)Akselrod, Argyropoulos, Hoang,
  Cirac{\`{\i}}, Fang, Huang, Smith, and Mikkelsen]{Akselrod2014a}
Akselrod,~G.~M.; Argyropoulos,~C.; Hoang,~T.~B.; Cirac{\`{\i}},~C.; Fang,~C.;
  Huang,~J.; Smith,~D.~R.; Mikkelsen,~M.~H. {Probing the mechanisms of large
  Purcell enhancement in plasmonic nanoantennas}. \emph{Nature Photonics}
  \textbf{2014}, \emph{8}, 835--840\relax
\mciteBstWouldAddEndPuncttrue
\mciteSetBstMidEndSepPunct{\mcitedefaultmidpunct}
{\mcitedefaultendpunct}{\mcitedefaultseppunct}\relax
\EndOfBibitem
\bibitem[Kinkhabwala \latin{et~al.}(2009)Kinkhabwala, Yu, Fan, Avlasevich,
  M{\"{u}}llen, and Moerner]{Kinkhabwala2009}
Kinkhabwala,~A.; Yu,~Z.; Fan,~S.; Avlasevich,~Y.; M{\"{u}}llen,~K.;
  Moerner,~W.~E. {Large single-molecule fluorescence enhancements produced by a
  bowtie nanoantenna}. \emph{Nature Photonics} \textbf{2009}, \emph{3},
  654--657\relax
\mciteBstWouldAddEndPuncttrue
\mciteSetBstMidEndSepPunct{\mcitedefaultmidpunct}
{\mcitedefaultendpunct}{\mcitedefaultseppunct}\relax
\EndOfBibitem
\bibitem[Pfeiffer \latin{et~al.}(2010)Pfeiffer, Lindfors, Wolpert, Atkinson,
  Benyoucef, Rastelli, Schmidt, Giessen, and Lippitz]{Pfeiffer2010}
Pfeiffer,~M.; Lindfors,~K.; Wolpert,~C.; Atkinson,~P.; Benyoucef,~M.;
  Rastelli,~A.; Schmidt,~O.~G.; Giessen,~H.; Lippitz,~M. {Enhancing the optical
  excitation efficiency of a single self-assembled quantum dot with a plasmonic
  nanoantenna.} \emph{Nano Letters} \textbf{2010}, \emph{10}, 4555--4558\relax
\mciteBstWouldAddEndPuncttrue
\mciteSetBstMidEndSepPunct{\mcitedefaultmidpunct}
{\mcitedefaultendpunct}{\mcitedefaultseppunct}\relax
\EndOfBibitem
\bibitem[Curto \latin{et~al.}(2010)Curto, Volpe, Taminiau, Kreuzer, Quidant,
  and van Hulst]{Curto:2010}
Curto,~A.~G.; Volpe,~G.; Taminiau,~T.~H.; Kreuzer,~M.~P.; Quidant,~R.; van
  Hulst,~N.~F. Unidirectional Emission of a Quantum Dot Coupled to a
  Nanoantenna. \emph{Science} \textbf{2010}, \emph{329}, 930--933\relax
\mciteBstWouldAddEndPuncttrue
\mciteSetBstMidEndSepPunct{\mcitedefaultmidpunct}
{\mcitedefaultendpunct}{\mcitedefaultseppunct}\relax
\EndOfBibitem
\bibitem[Dregely \latin{et~al.}(2014)Dregely, Lindfors, Lippitz, Engheta,
  Totzeck, and Giessen]{Dregely:2014}
Dregely,~D.; Lindfors,~K.; Lippitz,~M.; Engheta,~N.; Totzeck,~M.; Giessen,~H.
  Imaging and steering an optical wireless nanoantenna link. \emph{Nature
  Communications} \textbf{2014}, \emph{5}\relax
\mciteBstWouldAddEndPuncttrue
\mciteSetBstMidEndSepPunct{\mcitedefaultmidpunct}
{\mcitedefaultendpunct}{\mcitedefaultseppunct}\relax
\EndOfBibitem
\bibitem[Jun \latin{et~al.}(2011)Jun, Huang, and Brongersma]{Jun2011}
Jun,~Y.~C.; Huang,~K. C.~Y.; Brongersma,~M.~L. {Plasmonic beaming and active
  control over fluorescent emission.} \emph{Nature Communications}
  \textbf{2011}, \emph{2}, 283\relax
\mciteBstWouldAddEndPuncttrue
\mciteSetBstMidEndSepPunct{\mcitedefaultmidpunct}
{\mcitedefaultendpunct}{\mcitedefaultseppunct}\relax
\EndOfBibitem
\bibitem[Hartschuh \latin{et~al.}(2005)Hartschuh, Qian, Meixner, Anderson, and
  Novotny]{Hartschuh2005}
Hartschuh,~A.; Qian,~H.; Meixner,~A.~J.; Anderson,~N.; Novotny,~L. {Nanoscale
  optical imaging of excitons in single-walled carbon nanotubes}. \emph{Nano
  Letters} \textbf{2005}, \emph{5}, 2310--2313\relax
\mciteBstWouldAddEndPuncttrue
\mciteSetBstMidEndSepPunct{\mcitedefaultmidpunct}
{\mcitedefaultendpunct}{\mcitedefaultseppunct}\relax
\EndOfBibitem
\bibitem[Wokaun \latin{et~al.}(1983)Wokaun, Lutz, King, Wild, and
  Ernst]{Wokaun1983}
Wokaun,~A.; Lutz,~H.-P.; King,~A.~P.; Wild,~U.~P.; Ernst,~R.~R. {Energy
  transfer in surface enhanced luminescence}. \emph{The Journal of Chemical
  Physics} \textbf{1983}, \emph{79}, 509\relax
\mciteBstWouldAddEndPuncttrue
\mciteSetBstMidEndSepPunct{\mcitedefaultmidpunct}
{\mcitedefaultendpunct}{\mcitedefaultseppunct}\relax
\EndOfBibitem
\bibitem[St{\"{o}}ckle \latin{et~al.}(2000)St{\"{o}}ckle, Suh, Deckert, and
  Zenobi]{Stockle2000}
St{\"{o}}ckle,~R.~M.; Suh,~Y.~D.; Deckert,~V.; Zenobi,~R. {Nanoscale chemical
  analysis by tip-enhanced Raman spectroscopy}. \emph{Chemical Physics Letters}
  \textbf{2000}, \emph{318}, 131--136\relax
\mciteBstWouldAddEndPuncttrue
\mciteSetBstMidEndSepPunct{\mcitedefaultmidpunct}
{\mcitedefaultendpunct}{\mcitedefaultseppunct}\relax
\EndOfBibitem
\bibitem[Takase \latin{et~al.}(2013)Takase, Ajiki, Mizumoto, Komeda, Nara,
  Nabika, Yasuda, Ishihara, and Murakoshi]{Takase2013}
Takase,~M.; Ajiki,~H.; Mizumoto,~Y.; Komeda,~K.; Nara,~M.; Nabika,~H.;
  Yasuda,~S.; Ishihara,~H.; Murakoshi,~K. {Selection-rule breakdown in
  plasmon-induced electronic excitation of an isolated single-walled carbon
  nanotube}. \emph{Nature Photonics} \textbf{2013}, \emph{7}, 550--554\relax
\mciteBstWouldAddEndPuncttrue
\mciteSetBstMidEndSepPunct{\mcitedefaultmidpunct}
{\mcitedefaultendpunct}{\mcitedefaultseppunct}\relax
\EndOfBibitem
\bibitem[Jain \latin{et~al.}(2012)Jain, Ghosh, Baer, Rabani, and
  Alivisatos]{Jain2012}
Jain,~P.~K.; Ghosh,~D.; Baer,~R.; Rabani,~E.; Alivisatos,~A.~P. {Near-field
  manipulation of spectroscopic selection rules on the nanoscale.}
  \emph{Proceedings of the National Academy of Sciences of the United States of
  America} \textbf{2012}, \emph{109}, 8016--9\relax
\mciteBstWouldAddEndPuncttrue
\mciteSetBstMidEndSepPunct{\mcitedefaultmidpunct}
{\mcitedefaultendpunct}{\mcitedefaultseppunct}\relax
\EndOfBibitem
\bibitem[Iida and Ishihara(2009)Iida, and Ishihara]{Iida2009}
Iida,~T.; Ishihara,~H. {Unconventional control of excited states of a dimer
  molecule by a localized light field between metal nanostructures}.
  \emph{physica status solidi (a)} \textbf{2009}, \emph{206}, 980--984\relax
\mciteBstWouldAddEndPuncttrue
\mciteSetBstMidEndSepPunct{\mcitedefaultmidpunct}
{\mcitedefaultendpunct}{\mcitedefaultseppunct}\relax
\EndOfBibitem
\bibitem[Nakada \latin{et~al.}(1996)Nakada, Fujita, Dresselhaus, and
  Dresselhaus]{Nakada1996}
Nakada,~K.; Fujita,~M.; Dresselhaus,~G.; Dresselhaus,~M.~S. Edge state in
  graphene ribbons: Nanometer size effect and edge shape dependence.
  \emph{Phys. Rev. B} \textbf{1996}, \emph{54}, 17954--17961\relax
\mciteBstWouldAddEndPuncttrue
\mciteSetBstMidEndSepPunct{\mcitedefaultmidpunct}
{\mcitedefaultendpunct}{\mcitedefaultseppunct}\relax
\EndOfBibitem
\bibitem[Denk \latin{et~al.}(2014)Denk, Hohage, Zeppenfeld, Cai, Pignedoli,
  S{\"{o}}de, Fasel, Feng, M{\"{u}}llen, Wang, Prezzi, Ferretti, Ruini,
  Molinari, and Ruffieux]{Ruffieux_NatCom14}
Denk,~R.; Hohage,~M.; Zeppenfeld,~P.; Cai,~J.; Pignedoli,~C.~A.;
  S{\"{o}}de,~H.; Fasel,~R.; Feng,~X.; M{\"{u}}llen,~K.; Wang,~S.
  \latin{et~al.}  Exciton-dominated optical response of ultra-narrow graphene
  nanoribbons. \emph{Nature Communications} \textbf{2014}, \emph{5}, 4253\relax
\mciteBstWouldAddEndPuncttrue
\mciteSetBstMidEndSepPunct{\mcitedefaultmidpunct}
{\mcitedefaultendpunct}{\mcitedefaultseppunct}\relax
\EndOfBibitem
\bibitem[Chernov \latin{et~al.}(2013)Chernov, Fedotov, Talyzin, {Suarez Lopez},
  Anoshkin, Nasibulin, Kauppinen, and Obraztsova]{Chernov2013}
Chernov,~A.~I.; Fedotov,~P.~V.; Talyzin,~A.~V.; {Suarez Lopez},~I.;
  Anoshkin,~I.~V.; Nasibulin,~A.~G.; Kauppinen,~E.~I.; Obraztsova,~E.~D.
  {Optical properties of graphene nanoribbons encapsulated in single-walled
  carbon nanotubes}. \emph{ACS Nano} \textbf{2013}, \emph{7}, 6346--6353\relax
\mciteBstWouldAddEndPuncttrue
\mciteSetBstMidEndSepPunct{\mcitedefaultmidpunct}
{\mcitedefaultendpunct}{\mcitedefaultseppunct}\relax
\EndOfBibitem
\bibitem[Lim \latin{et~al.}(2015)Lim, Miyata, Fujihara, Okada, Liu, Arifin,
  Sato, Omachi, Kitaura, Irle, Suenaga, and Shinohara]{Lim2015}
Lim,~H.~E.; Miyata,~Y.; Fujihara,~M.; Okada,~S.; Liu,~Z.; Arifin,; Sato,~K.;
  Omachi,~H.; Kitaura,~R.; Irle,~S. \latin{et~al.}  {Fabrication and Optical
  Probing of Highly Extended, Ultrathin Graphene Nanoribbons in Carbon
  Nanotubes}. \emph{ACS Nano} \textbf{2015}, \emph{9}, 5034--5040\relax
\mciteBstWouldAddEndPuncttrue
\mciteSetBstMidEndSepPunct{\mcitedefaultmidpunct}
{\mcitedefaultendpunct}{\mcitedefaultseppunct}\relax
\EndOfBibitem
\bibitem[Li \latin{et~al.}(2007)Li, Marvin, and Steven]{louie07-excitonic}
Li,~Y.; Marvin,~L.~C.; Steven,~G.~L. Excitonic Effects in the Optical Spectra
  of Graphene Nanoribbons. \emph{Nano Letters} \textbf{2007}, \emph{7},
  3112--3115\relax
\mciteBstWouldAddEndPuncttrue
\mciteSetBstMidEndSepPunct{\mcitedefaultmidpunct}
{\mcitedefaultendpunct}{\mcitedefaultseppunct}\relax
\EndOfBibitem
\bibitem[Prezzi \latin{et~al.}(2008)Prezzi, Varsano, Ruini, Marini, and
  Molinari]{prezzi08-manybody}
Prezzi,~D.; Varsano,~D.; Ruini,~A.; Marini,~A.; Molinari,~E. Optical properties
  of graphene nanoribbons: The role of many-body effects. \emph{Phys. Rev. B}
  \textbf{2008}, \emph{77}, 041404\relax
\mciteBstWouldAddEndPuncttrue
\mciteSetBstMidEndSepPunct{\mcitedefaultmidpunct}
{\mcitedefaultendpunct}{\mcitedefaultseppunct}\relax
\EndOfBibitem
\bibitem[Kim and Kim(2013)Kim, and Kim]{Kim-NatNan13}
Kim,~W.~Y.; Kim,~K.~S. Prediction of very large values of magnetoresistance in
  a graphene nanoribbon device. \emph{Nature Nanotechnology} \textbf{2013},
  \emph{3}, 408--412\relax
\mciteBstWouldAddEndPuncttrue
\mciteSetBstMidEndSepPunct{\mcitedefaultmidpunct}
{\mcitedefaultendpunct}{\mcitedefaultseppunct}\relax
\EndOfBibitem
\bibitem[Stampfer \latin{et~al.}(2009)Stampfer, G\"ottinger, Hellm\"oller,
  Molitor, Ensslin, and Ihn]{Stampfer2009}
Stampfer,~C.; G\"ottinger,~J.; Hellm\"oller,~S.; Molitor,~F.; Ensslin,~K.;
  Ihn,~T. {Energy gaps in etched graphene nanoribbons}. \emph{Phys. Rev. Lett.}
  \textbf{2009}, \emph{102}, 056403\relax
\mciteBstWouldAddEndPuncttrue
\mciteSetBstMidEndSepPunct{\mcitedefaultmidpunct}
{\mcitedefaultendpunct}{\mcitedefaultseppunct}\relax
\EndOfBibitem
\bibitem[Li \latin{et~al.}(2008)Li, Wang, Zhang, Lee, and Dai]{Li2008}
Li,~X.; Wang,~X.; Zhang,~L.; Lee,~S.; Dai,~H. {Chemically Derived, Ultrasmooth
  Graphene Nanoribbon Semiconductors}. \emph{Science} \textbf{2008},
  \emph{319}, 1229--1232\relax
\mciteBstWouldAddEndPuncttrue
\mciteSetBstMidEndSepPunct{\mcitedefaultmidpunct}
{\mcitedefaultendpunct}{\mcitedefaultseppunct}\relax
\EndOfBibitem
\bibitem[Son \latin{et~al.}(2006)Son, Cohen, and Louie]{Son2006}
Son,~Y.-W.; Cohen,~M.~L.; Louie,~S.~G. {Half-Metallic Graphene Nanoribbons}.
  \emph{Nature} \textbf{2006}, \emph{444}, 347--349\relax
\mciteBstWouldAddEndPuncttrue
\mciteSetBstMidEndSepPunct{\mcitedefaultmidpunct}
{\mcitedefaultendpunct}{\mcitedefaultseppunct}\relax
\EndOfBibitem
\bibitem[Han \latin{et~al.}(2007)Han, Ozyilmaz, Zhang, and Kim]{Han2007}
Han,~M.~Y.; Ozyilmaz,~B.; Zhang,~Y.; Kim,~P. Energy Band-Gap Engineering of
  Graphene Nanoribbons. \emph{Phys. Rev. Lett.} \textbf{2007}, \emph{98},
  206805\relax
\mciteBstWouldAddEndPuncttrue
\mciteSetBstMidEndSepPunct{\mcitedefaultmidpunct}
{\mcitedefaultendpunct}{\mcitedefaultseppunct}\relax
\EndOfBibitem
\bibitem[Senkovskiy \latin{et~al.}(2017)Senkovskiy, Pfeiffer, Alavi, Bliesener,
  Zhu, Michel, Fedorov, German, Hertel, Haberer, Petaccia, Fischer, Meerholz,
  van Loosdrecht, Lindfors, and Gr\"uneis]{Senkovskiy17}
Senkovskiy,~B.~V.; Pfeiffer,~M.; Alavi,~S.~K.; Bliesener,~A.; Zhu,~J.;
  Michel,~S.; Fedorov,~A.~V.; German,~R.; Hertel,~D.; Haberer,~D.
  \latin{et~al.}  Making Graphene Nanoribbons Photoluminescent. \emph{Nano
  Letters} \textbf{2017}, \emph{17}, 4029--4037\relax
\mciteBstWouldAddEndPuncttrue
\mciteSetBstMidEndSepPunct{\mcitedefaultmidpunct}
{\mcitedefaultendpunct}{\mcitedefaultseppunct}\relax
\EndOfBibitem
\bibitem[Chen \latin{et~al.}(2013)Chen, de~Oteyza, Pedramrazi, Chen, Fischer,
  and Crommie]{Chen2013}
Chen,~Y.-C.; de~Oteyza,~D.~G.; Pedramrazi,~Z.; Chen,~C.; Fischer,~F.~R.;
  Crommie,~M.~F. {Tuning the Band Gap of Graphene Nanoribbons Synthesized from
  Molecular Precursors}. \emph{ACS Nano} \textbf{2013}, \emph{7},
  6123--6128\relax
\mciteBstWouldAddEndPuncttrue
\mciteSetBstMidEndSepPunct{\mcitedefaultmidpunct}
{\mcitedefaultendpunct}{\mcitedefaultseppunct}\relax
\EndOfBibitem
\bibitem[Cai \latin{et~al.}(2010)Cai, Ruffieux, Jaafar, Bieri, Braun,
  Blankenburg, Muoth, Seitsonen, Saleh, Feng, Mullen, and
  Fasel]{muellen10-ribbon}
Cai,~J.; Ruffieux,~P.; Jaafar,~R.; Bieri,~M.; Braun,~T.; Blankenburg,~S.;
  Muoth,~M.; Seitsonen,~A.~P.; Saleh,~M.; Feng,~X. \latin{et~al.}  Atomically
  precise bottom-up fabrication of graphene nanoribbons. \emph{Nature}
  \textbf{2010}, \emph{466}, 470--473\relax
\mciteBstWouldAddEndPuncttrue
\mciteSetBstMidEndSepPunct{\mcitedefaultmidpunct}
{\mcitedefaultendpunct}{\mcitedefaultseppunct}\relax
\EndOfBibitem
\bibitem[Ruffieux \latin{et~al.}(2016)Ruffieux, Wang, Yang,
  S\`{a}nchez-S\`{a}nchez, Liu, Dienel, Talirz, Shinde, Pignedoli, Passerone,
  Dumslaff, Feng, M\"{u}llen, and Fasel]{Ruffieux_ZigZag}
Ruffieux,~P.; Wang,~S.; Yang,~B.; S\`{a}nchez-S\`{a}nchez,~C.; Liu,~J.;
  Dienel,~T.; Talirz,~L.; Shinde,~P.; Pignedoli,~C.~A.; Passerone,~D.
  \latin{et~al.}  On-surface synthesis of graphene nanoribbons with zigzag edge
  topology. \emph{Nature} \textbf{2016}, \emph{531}, 489–492\relax
\mciteBstWouldAddEndPuncttrue
\mciteSetBstMidEndSepPunct{\mcitedefaultmidpunct}
{\mcitedefaultendpunct}{\mcitedefaultseppunct}\relax
\EndOfBibitem
\bibitem[Talirz \latin{et~al.}(2016)Talirz, Ruffieux, and Fasel]{ADMA201505738}
Talirz,~L.; Ruffieux,~P.; Fasel,~R. On-Surface Synthesis of Atomically Precise
  Graphene Nanoribbons. \emph{Advanced Materials} \textbf{2016}, \emph{28},
  6222--6231\relax
\mciteBstWouldAddEndPuncttrue
\mciteSetBstMidEndSepPunct{\mcitedefaultmidpunct}
{\mcitedefaultendpunct}{\mcitedefaultseppunct}\relax
\EndOfBibitem
\bibitem[Chen \latin{et~al.}(2015)Chen, Cao, Chen, Pedramrazi, Haberer,
  de~Oteyza, Fischer, Louie, and Crommie]{Chen15-heterojunction}
Chen,~Y.-C.; Cao,~T.; Chen,~C.; Pedramrazi,~Z.; Haberer,~D.; de~Oteyza,~D.;
  Fischer,~F.; Louie,~S.; Crommie,~M. Molecular bandgap engineering of
  bottom-up synthesized graphene nanoribbon heterojunctions. \emph{Nature
  Nanotechnology} \textbf{2015}, \emph{10}, 156–160\relax
\mciteBstWouldAddEndPuncttrue
\mciteSetBstMidEndSepPunct{\mcitedefaultmidpunct}
{\mcitedefaultendpunct}{\mcitedefaultseppunct}\relax
\EndOfBibitem
\bibitem[Kawai \latin{et~al.}(2015)Kawai, Saito, Osumi, Yamaguchi, Foster,
  Spijker, and Meyer]{B-GNRs}
Kawai,~S.; Saito,~S.; Osumi,~S.; Yamaguchi,~S.; Foster,~A.~S.; Spijker,~P.;
  Meyer,~E. Atomically controlled substitutional boron-doping of graphene
  nanoribbons. \emph{Nature Communications} \textbf{2015}, \emph{6}, 8098\relax
\mciteBstWouldAddEndPuncttrue
\mciteSetBstMidEndSepPunct{\mcitedefaultmidpunct}
{\mcitedefaultendpunct}{\mcitedefaultseppunct}\relax
\EndOfBibitem
\bibitem[Cloke \latin{et~al.}(2015)Cloke, Marangoni, Nguyen, Joshi, Rizzo,
  Bronner, Cao, Louie, Crommie, and Fischer]{Fischer_JACS_15}
Cloke,~R.~R.; Marangoni,~T.; Nguyen,~G.~D.; Joshi,~T.; Rizzo,~D.~J.;
  Bronner,~C.; Cao,~T.; Louie,~S.~G.; Crommie,~M.~F.; Fischer,~F.~R.
  Site-Specific Substitutional Boron Doping of Semiconducting Armchair Graphene
  Nanoribbons. \emph{Journal of the American Chemical Society} \textbf{2015},
  \emph{137}, 8872--8875\relax
\mciteBstWouldAddEndPuncttrue
\mciteSetBstMidEndSepPunct{\mcitedefaultmidpunct}
{\mcitedefaultendpunct}{\mcitedefaultseppunct}\relax
\EndOfBibitem
\bibitem[Carbonell-Sanrom{\`{a}} \latin{et~al.}(2017)Carbonell-Sanrom{\`{a}},
  Brandimarte, Balog, Corso, Kawai, Garcia-Lekue, Saito, Yamaguchi, Meyer,
  S{\'{a}}nchez-Portal, and Pascual]{Carbonell-Sanroma2017}
Carbonell-Sanrom{\`{a}},~E.; Brandimarte,~P.; Balog,~R.; Corso,~M.; Kawai,~S.;
  Garcia-Lekue,~A.; Saito,~S.; Yamaguchi,~S.; Meyer,~E.;
  S{\'{a}}nchez-Portal,~D. \latin{et~al.}  {Quantum Dots Embedded in Graphene
  Nanoribbons by Chemical Substitution}. \emph{Nano Letters} \textbf{2017},
  \emph{17}, 50--56\relax
\mciteBstWouldAddEndPuncttrue
\mciteSetBstMidEndSepPunct{\mcitedefaultmidpunct}
{\mcitedefaultendpunct}{\mcitedefaultseppunct}\relax
\EndOfBibitem
\bibitem[Nguyen \latin{et~al.}(2016)Nguyen, Toma, Cao, Pedramrazi, Chen, Rizzo,
  Joshi, Bronner, Chen, Favaro, Louie, Fischer, and Crommie]{Nguyen2016}
Nguyen,~G.~D.; Toma,~F.~M.; Cao,~T.; Pedramrazi,~Z.; Chen,~C.; Rizzo,~D.~J.;
  Joshi,~T.; Bronner,~C.; Chen,~Y.~C.; Favaro,~M. \latin{et~al.}  {Bottom-Up
  Synthesis of N = 13 Sulfur-Doped Graphene Nanoribbons}. \emph{Journal of
  Physical Chemistry C} \textbf{2016}, \emph{120}, 2684--2687\relax
\mciteBstWouldAddEndPuncttrue
\mciteSetBstMidEndSepPunct{\mcitedefaultmidpunct}
{\mcitedefaultendpunct}{\mcitedefaultseppunct}\relax
\EndOfBibitem
\bibitem[Wang \latin{et~al.}(2016)Wang, Zhou, Li, Li, Wu, Duan, and
  He]{Wang2016}
Wang,~W.-X.; Zhou,~M.; Li,~X.; Li,~S.-Y.; Wu,~X.; Duan,~W.; He,~L. {Energy gaps
  of atomically precise armchair graphene sidewall nanoribbons}. \emph{Phys.
  Rev. B} \textbf{2016}, \emph{93}, 241403\relax
\mciteBstWouldAddEndPuncttrue
\mciteSetBstMidEndSepPunct{\mcitedefaultmidpunct}
{\mcitedefaultendpunct}{\mcitedefaultseppunct}\relax
\EndOfBibitem
\bibitem[Chong \latin{et~al.}(0)Chong, Afsharimani, Scheurer, Cardoso,
  Ferretti, Prezzi, and Schull]{Chong2017}
Chong,~M.~C.; Afsharimani,~N.; Scheurer,~F.; Cardoso,~C.; Ferretti,~A.;
  Prezzi,~D.; Schull,~G. Bright electroluminescence from single graphene
  nanoribbon junctions. \emph{Nano Letters} \textbf{0}, \emph{0}, null\relax
\mciteBstWouldAddEndPuncttrue
\mciteSetBstMidEndSepPunct{\mcitedefaultmidpunct}
{\mcitedefaultendpunct}{\mcitedefaultseppunct}\relax
\EndOfBibitem
\bibitem[Zhao \latin{et~al.}(2017)Zhao, Borin~Barin, Rondin, Raynaud,
  Fairbrother, Dumslaff, Campidelli, Müllen, Narita, Voisin, Ruffieux, Fasel,
  and Lauret]{Zhao2017}
Zhao,~S.; Borin~Barin,~G.; Rondin,~L.; Raynaud,~C.; Fairbrother,~A.;
  Dumslaff,~T.; Campidelli,~S.; Müllen,~K.; Narita,~A.; Voisin,~C.
  \latin{et~al.}  Optical Investigation of On-Surface Synthesized Armchair
  Graphene Nanoribbons. \emph{physica status solidi (b)} \textbf{2017},
  \emph{254}, 1700223\relax
\mciteBstWouldAddEndPuncttrue
\mciteSetBstMidEndSepPunct{\mcitedefaultmidpunct}
{\mcitedefaultendpunct}{\mcitedefaultseppunct}\relax
\EndOfBibitem
\bibitem[Yang \latin{et~al.}(2007)Yang, Cohen, and Louie]{Yang2007}
Yang,~L.; Cohen,~M.~L.; Louie,~S.~G. {Excitonic effects in the optical spectra
  of graphene nanoribbons}. \emph{Nano Letters} \textbf{2007}, \emph{7},
  3112--3115\relax
\mciteBstWouldAddEndPuncttrue
\mciteSetBstMidEndSepPunct{\mcitedefaultmidpunct}
{\mcitedefaultendpunct}{\mcitedefaultseppunct}\relax
\EndOfBibitem
\bibitem[Prezzi \latin{et~al.}(2007)Prezzi, Varsano, Ruini, Marini, and
  Molinari]{Prezzi2007}
Prezzi,~D.; Varsano,~D.; Ruini,~A.; Marini,~A.; Molinari,~E. {Optical
  properties of graphene nanoribbons: the role of many-body effects}.
  \emph{Phys. Rev. B} \textbf{2007}, \emph{77}, 041404\relax
\mciteBstWouldAddEndPuncttrue
\mciteSetBstMidEndSepPunct{\mcitedefaultmidpunct}
{\mcitedefaultendpunct}{\mcitedefaultseppunct}\relax
\EndOfBibitem
\bibitem[Zhao \latin{et~al.}(2017)Zhao, Rondin, Delport, Voisin, Beser, Hu,
  Feng, MÃ¼llen, Narita, Campidelli, and Lauret]{Zhao2017_SciDir}
Zhao,~S.; Rondin,~L.; Delport,~G.; Voisin,~C.; Beser,~U.; Hu,~Y.; Feng,~X.;
  MÃ¼llen,~K.; Narita,~A.; Campidelli,~S. \latin{et~al.}  Fluorescence from
  graphene nanoribbons of well-defined structure. \emph{Carbon} \textbf{2017},
  \emph{119}, 235 -- 240\relax
\mciteBstWouldAddEndPuncttrue
\mciteSetBstMidEndSepPunct{\mcitedefaultmidpunct}
{\mcitedefaultendpunct}{\mcitedefaultseppunct}\relax
\EndOfBibitem
\bibitem[Gao \latin{et~al.}(2012)Gao, Ren, Xu, Jin, Wang, Ma, Ma, Zhang, Fu,
  Peng, Bao, and Cheng]{Gao2012}
Gao,~L.; Ren,~W.; Xu,~H.; Jin,~L.; Wang,~Z.; Ma,~T.; Ma,~L.-P.; Zhang,~Z.;
  Fu,~Q.; Peng,~L.-M. \latin{et~al.}  {Repeated growth and bubbling transfer of
  graphene with millimetre-size single-crystal grains using platinum}.
  \emph{Nature Communications} \textbf{2012}, \emph{3}, 699\relax
\mciteBstWouldAddEndPuncttrue
\mciteSetBstMidEndSepPunct{\mcitedefaultmidpunct}
{\mcitedefaultendpunct}{\mcitedefaultseppunct}\relax
\EndOfBibitem
\bibitem[Wang \latin{et~al.}(2011)Wang, Stobbe, and Lodahl]{Wang2011}
Wang,~Q.; Stobbe,~S.; Lodahl,~P. Mapping the local density of states of a
  photonic crystal with single quantum dots. \emph{Phys. Rev. Lett.}
  \textbf{2011}, \emph{107}, 167404\relax
\mciteBstWouldAddEndPuncttrue
\mciteSetBstMidEndSepPunct{\mcitedefaultmidpunct}
{\mcitedefaultendpunct}{\mcitedefaultseppunct}\relax
\EndOfBibitem
\bibitem[Verzhbitskiy \latin{et~al.}(2016)Verzhbitskiy, Corato, Ruini,
  Molinari, Narita, Hu, Schwab, Bruna, Yoon, Milana, Feng, MÃ¼llen, Ferrari,
  Casiraghi, and Prezzi]{Verzhbitskiy16}
Verzhbitskiy,~I.~A.; Corato,~M.~D.; Ruini,~A.; Molinari,~E.; Narita,~A.;
  Hu,~Y.; Schwab,~M.~G.; Bruna,~M.; Yoon,~D.; Milana,~S. \latin{et~al.}  Raman
  Fingerprints of Atomically Precise Graphene Nanoribbons. \emph{Nano Letters}
  \textbf{2016}, \emph{16}, 3442--3447\relax
\mciteBstWouldAddEndPuncttrue
\mciteSetBstMidEndSepPunct{\mcitedefaultmidpunct}
{\mcitedefaultendpunct}{\mcitedefaultseppunct}\relax
\EndOfBibitem
\bibitem[Narita \latin{et~al.}(2014)Narita, Feng, Hernandez, Jensen, Bonn,
  Yang, Verzhbitskiy, Casiraghi, Hansen, Koch, Fytas, Ivasenko, Li, Mali,
  Balandina, Mahesh, De~Feyter, and Müllen]{Narita2014}
Narita,~A.; Feng,~X.; Hernandez,~Y.; Jensen,~S.~A.; Bonn,~M.; Yang,~H.;
  Verzhbitskiy,~I.~A.; Casiraghi,~C.; Hansen,~M.~R.; Koch,~A. H.~R.
  \latin{et~al.}  Synthesis of structurally well-defined and
  liquid-phase-processable graphene nanoribbons. \emph{Nature Chemistry}
  \textbf{2014}, \emph{6}, 126--132\relax
\mciteBstWouldAddEndPuncttrue
\mciteSetBstMidEndSepPunct{\mcitedefaultmidpunct}
{\mcitedefaultendpunct}{\mcitedefaultseppunct}\relax
\EndOfBibitem
\bibitem[Gillen \latin{et~al.}(2009)Gillen, Mohr, Thomsen, and
  Maultzsch]{Gillen2009}
Gillen,~R.; Mohr,~M.; Thomsen,~C.; Maultzsch,~J. Vibrational properties of
  graphene nanoribbons by first-principles calculations. \emph{Phys. Rev. B}
  \textbf{2009}, \emph{80}, 155418\relax
\mciteBstWouldAddEndPuncttrue
\mciteSetBstMidEndSepPunct{\mcitedefaultmidpunct}
{\mcitedefaultendpunct}{\mcitedefaultseppunct}\relax
\EndOfBibitem
\bibitem[Gillen \latin{et~al.}(2010)Gillen, Mohr, and Maultzsch]{Gillen2010}
Gillen,~R.; Mohr,~M.; Maultzsch,~J. Raman-active modes in graphene nanoribbons.
  \emph{physica status solidi (b)} \textbf{2010}, \emph{247}, 2941--2944\relax
\mciteBstWouldAddEndPuncttrue
\mciteSetBstMidEndSepPunct{\mcitedefaultmidpunct}
{\mcitedefaultendpunct}{\mcitedefaultseppunct}\relax
\EndOfBibitem
\bibitem[Ma \latin{et~al.}(2017)Ma, Liang, Xiao, Puretzky, Hong, Lu, Meunier,
  Bernholc, and Li]{Ma2017}
Ma,~C.; Liang,~L.; Xiao,~Z.; Puretzky,~A.~A.; Hong,~K.; Lu,~W.; Meunier,~V.;
  Bernholc,~J.; Li,~A.-P. Seamless Staircase Electrical Contact to
  Semiconducting Graphene Nanoribbons. \emph{Nano Letters} \textbf{2017},
  \emph{17}, 6241--6247\relax
\mciteBstWouldAddEndPuncttrue
\mciteSetBstMidEndSepPunct{\mcitedefaultmidpunct}
{\mcitedefaultendpunct}{\mcitedefaultseppunct}\relax
\EndOfBibitem
\bibitem[Can\ifmmode~\mbox{\c{c}}\else \c{c}\fi{}ado
  \latin{et~al.}(2014)Can\ifmmode~\mbox{\c{c}}\else \c{c}\fi{}ado, Beams,
  Jorio, and Novotny]{Cancado2014}
Can\ifmmode~\mbox{\c{c}}\else \c{c}\fi{}ado,~L.~G.; Beams,~R.; Jorio,~A.;
  Novotny,~L. Theory of Spatial Coherence in Near-Field Raman Scattering.
  \emph{Phys. Rev. X} \textbf{2014}, \emph{4}, 031054\relax
\mciteBstWouldAddEndPuncttrue
\mciteSetBstMidEndSepPunct{\mcitedefaultmidpunct}
{\mcitedefaultendpunct}{\mcitedefaultseppunct}\relax
\EndOfBibitem
\bibitem[Beams \latin{et~al.}(2014)Beams, Can\ifmmode~\mbox{\c{c}}\else
  \c{c}\fi{}ado, Oh, Jorio, and Novotny]{Beams2014}
Beams,~R.; Can\ifmmode~\mbox{\c{c}}\else \c{c}\fi{}ado,~L.~G.; Oh,~S.-H.;
  Jorio,~A.; Novotny,~L. Spatial Coherence in Near-Field Raman Scattering.
  \emph{Phys. Rev. Lett.} \textbf{2014}, \emph{113}, 186101\relax
\mciteBstWouldAddEndPuncttrue
\mciteSetBstMidEndSepPunct{\mcitedefaultmidpunct}
{\mcitedefaultendpunct}{\mcitedefaultseppunct}\relax
\EndOfBibitem
\bibitem[Messinger \latin{et~al.}(1981)Messinger, von Raben, Chang, and
  Barber]{Messinger1981}
Messinger,~B.~J.; von Raben,~K.~U.; Chang,~R.~K.; Barber,~P.~W. {Local fields
  at the surface of noble-metal microspheres}. \emph{Phys. Rev. B}
  \textbf{1981}, \emph{24}, 649--657\relax
\mciteBstWouldAddEndPuncttrue
\mciteSetBstMidEndSepPunct{\mcitedefaultmidpunct}
{\mcitedefaultendpunct}{\mcitedefaultseppunct}\relax
\EndOfBibitem
\bibitem[Alonso-Gonz{\'{a}}lez \latin{et~al.}(2013)Alonso-Gonz{\'{a}}lez,
  Albella, Neubrech, Huck, Chen, Golmar, Casanova, Hueso, Pucci, Aizpurua, and
  Hillenbrand]{Alonso-Gonzalez2013}
Alonso-Gonz{\'{a}}lez,~P.; Albella,~P.; Neubrech,~F.; Huck,~C.; Chen,~J.;
  Golmar,~F.; Casanova,~F.; Hueso,~L.~E.; Pucci,~A.; Aizpurua,~J.
  \latin{et~al.}  {Experimental Verification of the Spectral Shift between
  Near- and Far-Field Peak Intensities of Plasmonic Infrared Nanoantennas}.
  \emph{Phys. Rev Lett.} \textbf{2013}, \emph{110}, 203902\relax
\mciteBstWouldAddEndPuncttrue
\mciteSetBstMidEndSepPunct{\mcitedefaultmidpunct}
{\mcitedefaultendpunct}{\mcitedefaultseppunct}\relax
\EndOfBibitem
\bibitem[Zuloaga and Nordlander(2011)Zuloaga, and Nordlander]{Zuloaga2011}
Zuloaga,~J.; Nordlander,~P. {On the Energy Shift between Near-Field and
  Far-Field Peak Intensities in Localized Plasmon Systems}. \emph{Nano Letters}
  \textbf{2011}, \emph{11}, 1280--1283\relax
\mciteBstWouldAddEndPuncttrue
\mciteSetBstMidEndSepPunct{\mcitedefaultmidpunct}
{\mcitedefaultendpunct}{\mcitedefaultseppunct}\relax
\EndOfBibitem
\bibitem[Hecht and Novotny(2006)Hecht, and Novotny]{Hecht2006}
Hecht,~B.; Novotny,~L. \emph{Principles of nanooptics}; Cambridge University
  Press, 2006\relax
\mciteBstWouldAddEndPuncttrue
\mciteSetBstMidEndSepPunct{\mcitedefaultmidpunct}
{\mcitedefaultendpunct}{\mcitedefaultseppunct}\relax
\EndOfBibitem
\bibitem[Janssen \latin{et~al.}(2010)Janssen, Wachters, and
  Urbach]{Janssen2010}
Janssen,~O. T.~A.; Wachters,~A. J.~H.; Urbach,~H.~P. Efficient optimization
  method for the light extraction from periodically modulated LEDs using
  reciprocity. \emph{Opt. Express} \textbf{2010}, \emph{18}, 24522--24535\relax
\mciteBstWouldAddEndPuncttrue
\mciteSetBstMidEndSepPunct{\mcitedefaultmidpunct}
{\mcitedefaultendpunct}{\mcitedefaultseppunct}\relax
\EndOfBibitem
\bibitem[Zhang \latin{et~al.}(2015)Zhang, Martins, Diyaf, Wilson, Turnbull, and
  Samuel]{Zhang2015}
Zhang,~S.; Martins,~E.~R.; Diyaf,~A.~G.; Wilson,~J.~I.; Turnbull,~G.~A.;
  Samuel,~I.~D. Calculation of the emission power distribution of
  microstructured \{OLEDs\} using the reciprocity theorem. \emph{Synthetic
  Metals} \textbf{2015}, \emph{205}, 127 -- 133\relax
\mciteBstWouldAddEndPuncttrue
\mciteSetBstMidEndSepPunct{\mcitedefaultmidpunct}
{\mcitedefaultendpunct}{\mcitedefaultseppunct}\relax
\EndOfBibitem
\bibitem[Chen and Koenderink(2015)Chen, and Koenderink]{Chen:15}
Chen,~Y.; Koenderink,~A.~F. General point dipole theory for periodic
  metasurfaces: magnetoelectric scattering lattices coupled to planar photonic
  structures. \emph{arXiv:1512.00582} \textbf{2015}, \relax
\mciteBstWouldAddEndPunctfalse
\mciteSetBstMidEndSepPunct{\mcitedefaultmidpunct}
{}{\mcitedefaultseppunct}\relax
\EndOfBibitem
\bibitem[Pfeiffer \latin{et~al.}(2014)Pfeiffer, Lindfors, Zhang, Fenk,
  Phillipp, Atkinson, Rastelli, Schmidt, Giessen, and Lippitz]{Pfeiffer2014}
Pfeiffer,~M.; Lindfors,~K.; Zhang,~H.; Fenk,~B.; Phillipp,~F.; Atkinson,~P.;
  Rastelli,~A.; Schmidt,~O.~G.; Giessen,~H.; Lippitz,~M. {Eleven nanometer
  alignment precision of a plasmonic nanoantenna with a self-assembled GaAs
  quantum dot.} \emph{Nano Letters} \textbf{2014}, \emph{14}, 197--201\relax
\mciteBstWouldAddEndPuncttrue
\mciteSetBstMidEndSepPunct{\mcitedefaultmidpunct}
{\mcitedefaultendpunct}{\mcitedefaultseppunct}\relax
\EndOfBibitem
\bibitem[Chen \latin{et~al.}(2017)Chen, Zhang, and Koenderink]{Koenderink17}
Chen,~Y.; Zhang,~Y.; Koenderink,~A.~F. General point dipole theory for periodic
  metasurfaces: magnetoelectric scattering lattices coupled to planar photonic
  structures. \emph{Opt. Express} \textbf{2017}, \emph{25}, 21358--21378\relax
\mciteBstWouldAddEndPuncttrue
\mciteSetBstMidEndSepPunct{\mcitedefaultmidpunct}
{\mcitedefaultendpunct}{\mcitedefaultseppunct}\relax
\EndOfBibitem
\bibitem[Thyagarajan \latin{et~al.}(2012)Thyagarajan, Rivier, Lovera, and
  Martin]{Thyagarajan2012}
Thyagarajan,~K.; Rivier,~S.; Lovera,~A.; Martin,~O.~J. {Enhanced
  second-harmonic generation from double resonant plasmonic antennae}.
  \emph{Optics Express} \textbf{2012}, \emph{20}, 12860\relax
\mciteBstWouldAddEndPuncttrue
\mciteSetBstMidEndSepPunct{\mcitedefaultmidpunct}
{\mcitedefaultendpunct}{\mcitedefaultseppunct}\relax
\EndOfBibitem
\bibitem[F{\'e}lidj \latin{et~al.}(2005)F{\'e}lidj, Laurent, Aubard, L{\'e}vi,
  Hohenau, Krenn, and Aussenegg]{Felidj2005}
F{\'e}lidj,~N.; Laurent,~G.; Aubard,~J.; L{\'e}vi,~G.; Hohenau,~A.;
  Krenn,~J.~R.; Aussenegg,~F.~R. Grating-induced plasmon mode in gold
  nanoparticle arrays. \emph{The Journal of Chemical Physics} \textbf{2005},
  \emph{123}, 221103\relax
\mciteBstWouldAddEndPuncttrue
\mciteSetBstMidEndSepPunct{\mcitedefaultmidpunct}
{\mcitedefaultendpunct}{\mcitedefaultseppunct}\relax
\EndOfBibitem
\bibitem[Christ(2005)]{Christ2005}
Christ,~A. Optical properties of metallic photonic crystal structures. Ph.D.\
  thesis, Philipps-Universit\"at Marburg, 2005\relax
\mciteBstWouldAddEndPuncttrue
\mciteSetBstMidEndSepPunct{\mcitedefaultmidpunct}
{\mcitedefaultendpunct}{\mcitedefaultseppunct}\relax
\EndOfBibitem
\bibitem[Lamprecht \latin{et~al.}(2000)Lamprecht, Schider, Lechner,
  Dietlbacher, Krenn, Leitner, and Aussenegg]{Lamprecht2000}
Lamprecht,~B.; Schider,~G.; Lechner,~R.~T.; Dietlbacher,~H.; Krenn,~J.~R.;
  Leitner,~A.; Aussenegg,~F.~R. Metal nanoparticle gratings: Influence of
  dipolar particle interaction on the plasmon resonance. \emph{Phys. Rev.
  Lett.} \textbf{2000}, \emph{84}, 4721\relax
\mciteBstWouldAddEndPuncttrue
\mciteSetBstMidEndSepPunct{\mcitedefaultmidpunct}
{\mcitedefaultendpunct}{\mcitedefaultseppunct}\relax
\EndOfBibitem
\bibitem[Auguié \latin{et~al.}(2010)Auguié, Benda$\tilde{n}$a, Barnes, and
  García~de Abajo]{Auguie2010}
Auguié,~B.; Benda$\tilde{n}$a,~X.~M.; Barnes,~W.~L.; García~de Abajo,~F.~J.
  Diffractive arrays of gold nanoparticles near an interface: Critical role of
  the substrate. \emph{Phys. Rev. B} \textbf{2010}, \emph{82}, 155447\relax
\mciteBstWouldAddEndPuncttrue
\mciteSetBstMidEndSepPunct{\mcitedefaultmidpunct}
{\mcitedefaultendpunct}{\mcitedefaultseppunct}\relax
\EndOfBibitem
\bibitem[Hicks \latin{et~al.}(2005)Hicks, Zou, Schatz, Spears, Van~Duyne,
  Gunnarsson, Rindzevicius, Kasemo, and K\"all]{Hicks2005}
Hicks,~E.~M.; Zou,~S.; Schatz,~G.~C.; Spears,~K.~G.; Van~Duyne,~R.~P.;
  Gunnarsson,~L.; Rindzevicius,~T.; Kasemo,~B.; K\"all,~M. Controlling Plasmon
  Line Shapes through Diffractive Coupling in Linear Arrays of Cylindrical
  Nanoparticles Fabricated by Electron Beam Lithography. \emph{Nano Letters}
  \textbf{2005}, \emph{5}, 1065--1070\relax
\mciteBstWouldAddEndPuncttrue
\mciteSetBstMidEndSepPunct{\mcitedefaultmidpunct}
{\mcitedefaultendpunct}{\mcitedefaultseppunct}\relax
\EndOfBibitem
\end{mcitethebibliography}

%

\begin{suppinfo}

\renewcommand{\thefigure}{S\arabic{figure}}

\setcounter{figure}{0}
\subsection{Details on finite element simulations}
The simple point dipole model for the relationship between measured far-field extinction and near-field enhancement can be improved by taking also the excitation process into account. If $I_{enh}(\lambda)$ is the collected emission at the wavelength $\lambda$, then the following relationship between the collected intensity and the incident intensity $I_0$ holds 
\begin{equation}
I_{enh}(\lambda) =  \left\{\left[I_{ext}(\lambda) k(\lambda)^{-4} \right]_{norm} \cdot f_{em}(\lambda) \right\} ~f_{ex}(\lambda_{ex})~  I_0,
\end{equation}
where the term in square brackets is the with the $k^{-4}$-factor corrected extinction spectrum, which has been normalized, and $f_{em}$ and $f_{ex}$ are the spectral enhancement factors for the emission and excitation processes, respectively. From the measurement we extract a maximum total enhancement $f_{tot,max} = f_{em,max} f_{ex} = 16$.

For deducing the brightness increase for the emission to the far-field when periodic boundary conditions are applied, it suffices to consider the interaction in one unit cell. From the quasi-periodic solution $\mathbf{E}_{solution}(x,y,z)$ for the electric field in the structure for an incident plane wave we obtain at the limit of $r_D \rightarrow \infty$ as intensity in the far-field
\begin{equation}
\left| \mathbf{E}(\mathbf{r}_D)\cdot\hat{\sigma}~\right|^2=~\left| \mathbf{E}_{solution}(x,y,z) \cdot \mathbf{\mu}_E \right|^2,
\end{equation}
where $\mathbf{r}_D$ is the position of the detector and $\hat{\sigma}$ defines the polarization of the detected emission of the emitter dipole $\mathbf{\mu}_E$.

The enhancement spectra for the array are evaluated in a 32~nm thick layer above the interface and around the array elements [red domain in the inset of Fig.~\ref{fig:fig2}b)] as
\begin{multline}
I_{y,array} ~\propto~  \frac{1}{V_{domain}} [  \int_{-px/2}^{px/2}\int_{-py/2}^{py/2}\int_{0}^{z_0} \left| E_{solution,y}(x,y,z)\right|^2 ~d^3r-\\ \int_{antenna} \left|E_{solution,y}(x,y,z)\right|^2 ~d^3r],
\end{multline}
where $V_{domain}$ is the volume of the investigated domain. The enhancement spectra are evaluated by normalizing to the intensity average for the same geometry without the array.

We further deduce the position resolved interaction of array and GNRs from our simulations. In the simulations we calculate the fractional radiative local density of optical states~\cite{Koenderink17} in a plane, which is situated at $z_0=32$~nm above the interface, and thus 2~nm above the array elements. In the inset of Fig.~\ref{fig:fig2}b) the green line in the cross section indicates the location.



\begin{figure}[t!]
	\centering
		\includegraphics[width=16cm]{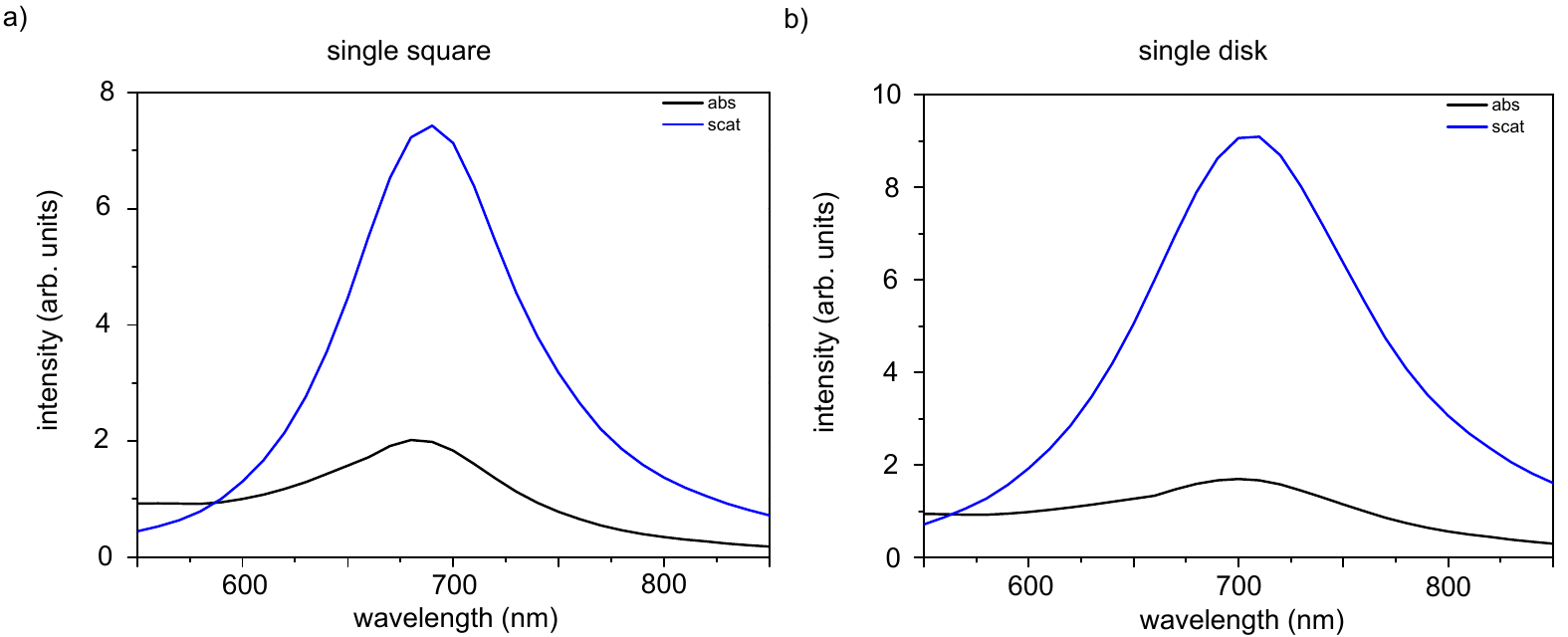}
	\caption{Absorption and scattering spectra for single isolated array elements: a) rectangle and b) round disk. The scattering clearly dominates over absorption.  }\label{fig:figS1}
\end{figure}

\subsection{Optical properties of single array elements}
Here we calculate the optical properties of the elements of which the arrays are composed of. The scattering and absorption spectra for the square and disk shaped elements are shown in Fig.~\ref{fig:figS1}a) and b), respectively. For these large element dimensions, the scattering clearly dominates the absorption. The single particle resonances are at 691~nm and 707~nm. The resonance widths are 102~nm and 135~nm.

\begin{figure}[t!]
	\centering
		\includegraphics[width=16cm]{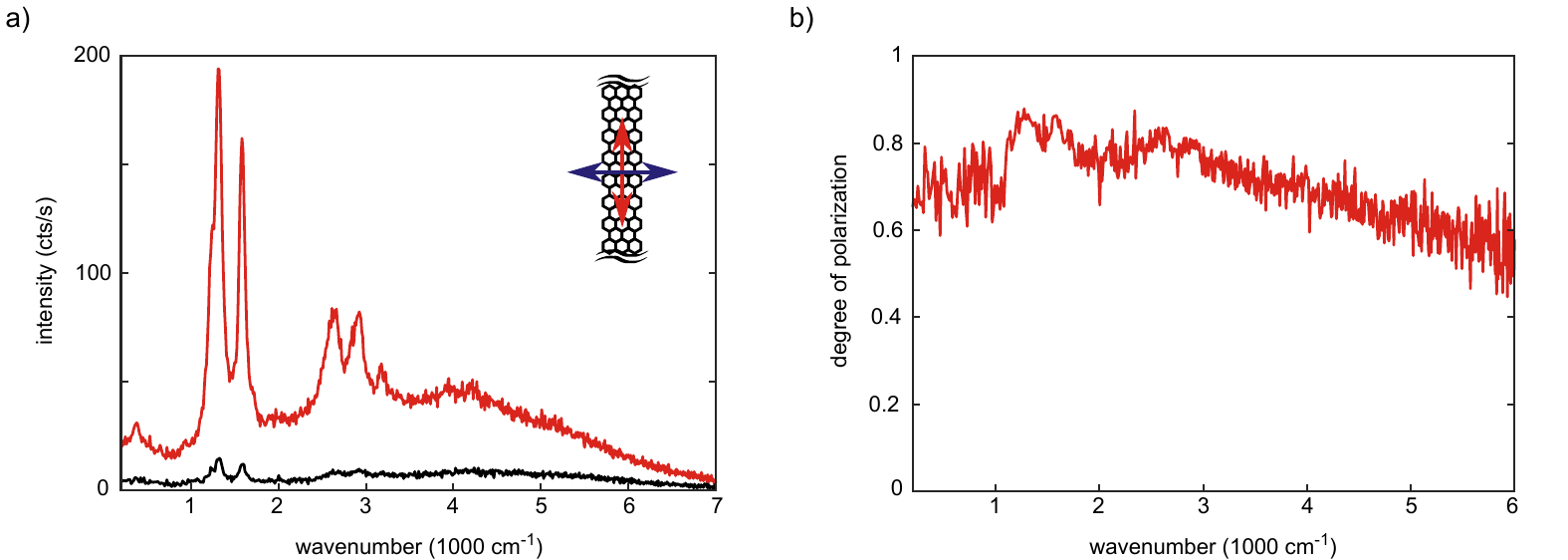}
	\caption{a) Spectra for polarization along the GNRs (red) and perpendicular (black) for 532~nm and corresponding degree of linear polarization b). }\label{fig:figS2}
\end{figure}

\subsection{Characterization of transfered GNRs}
The alignment of the GNRs is confirmed by measuring the photoluminescence and Raman spectra for polarization along and perpendicular to the orientation. The spectra for a laser wavelength of 532.8~nm are shown in Fig.~\ref{fig:figS2}a. The corresponding degree of linear polarization as a function of the relative shift is shown in panel b. Over almost the whole displayed frequency range, values of larger than 0.6 are obtained. By comparison with previous studies~\cite{Senkovskiy17}, this confirms the successful transfer of oriented 7-AGNRs.

\begin{figure}[t!]
	\centering
		\includegraphics[width=16cm]{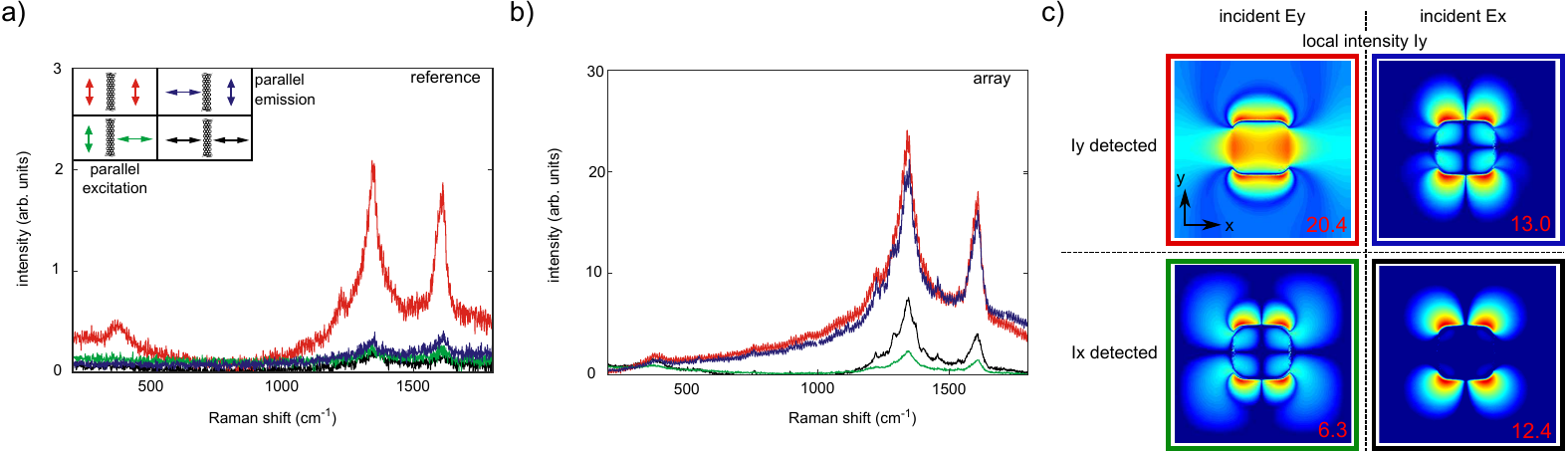}
	\caption{ Raman spectra for different combinations of excitation and detected polarizations for a laser wavelength of 633~nm. a) The reference spectra for GNRs on the substrate show a high polarization anisotropy. b) On the plasmonic array, Raman scattering is strongly enhanced for light scattered along the direction of the GNR orientation (blue). c) The trend for this observation is explained by the enhancement of the local intensity for orthogonal field components close to the antenna. 
    }\label{fig:figS3}
\end{figure}

\subsection{Local polarization enhancement close to plasmonic antennas}
To investigate the polarization dependent Raman spectra in Fig.~1, we compare polarization resolved spectral measurements with finite element simulations, which take the polarization of incident, local, and scattered/emitted fields into account. \\
In Fig.~\ref{fig:figS3}a the Raman scattering from GNRs is shown for each combination of polarizer and analyzer polarizer orientation along and perpendicular to the orientation direction of the GNRs. The largest signal intensity is obtained when polarizer and analyzer are oriented along the GNRs (red). The intensities for all other combinations are similarly weak, reproducing the results from Ref.~\cite{Senkovskiy17}.
For the GNRs on array (see Fig.~\ref{fig:figS3}b), both signals for analyzer polarizer parallel to the GNR orientation direction, independent of the incident polarization are of similar magnitude (red, blue) with the signal for parallel and parallel orientations being slightly stronger. The intensity for polarizer and analyzer both perpendicular to the GNR orientation (black) is here also stronger than for the spectrum (green) when the GNRs are illuminated with polarization along the GNR orientation and the perpendicular polarization signal is detected.


In analogy to the details on finite element simulations above, we calculate the intensity distributions around plasmonic antennas for both polarizations for incident plane waves at the wavelength of the incident laser and the scattered light. For the scattered light we assume interaction of optical phonons of the D-band in GNRs. From reciprocity considerations we obtain the full enhancement as the incoherent product of the enhancement patterns for excitation enhancement and emission enhancement.
We obtain the spatially resolved enhancement patterns in Fig.~\ref{fig:figS3}c), for the indicated incident and detected polarizations.
From the spatial average over the unit cell we deduce the enhancement factors (red numbers in lower right corner of each panel) corresponding to the spectra in Fig.~\ref{fig:figS3}b). 
The obtained values reproduce the experimentally observed trend, although, the shape and curvature of the antenna elements has a strong impact on the field localization and orientation, which becomes even more critical for the incoherent product of the intensity patterns for incident and scattered light. As we are lacking these structural informations, the simulations for the assumed geometry is thus not sufficiently robust for a quantitative comparison with the experimental data.

\begin{figure}[t!]
	\centering
		\includegraphics[width=16cm]{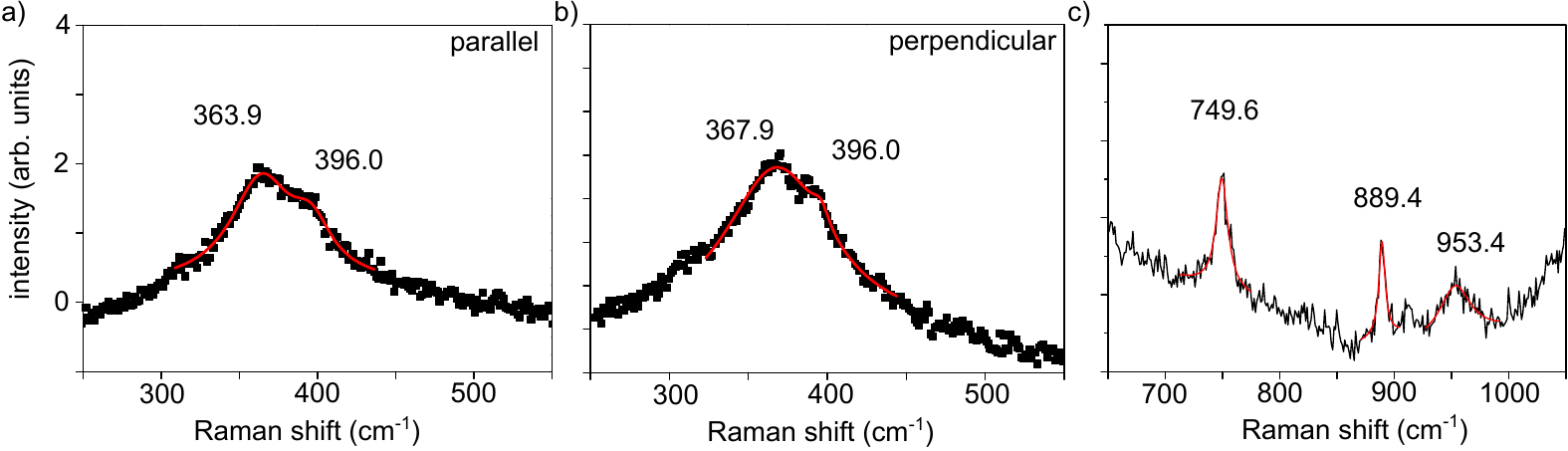}
	\caption{ Raman spectra for GNRs on array for parallel a) and perpendicular polarization b). The Raman shifts for the fitted peaks are indicated in the plots. c) Fit for the C-H vibrations and the 3-RBLM mode. 
    }\label{fig:figS4}
\end{figure}

\subsection{Analysis of the Raman modes}
In Fig.~\ref{fig:figS4}a and b we show the fitted peaks to the RBLM modes for GNRs on the array, for parallel and perpendicular polarization with respect to the GNR orientation direction, respectively. For the fit, we fix the Raman shift for the RBLM of pristine GNRs at the Raman shift of 396~cm$^{-1}$. The additionally determined Raman modes, which are attributed to C-H vibrational modes are displayed in Fig.~\ref{fig:figS4}c, together with the 3-RBLM mode. All fit functions are approximated as Lorentzians.


\end{suppinfo}

\end{document}